\documentclass[%
 reprint,
superscriptaddress,
 amsmath,amssymb,
 aps,
]{revtex4-2}

\usepackage{graphicx}
\usepackage{dcolumn}
\usepackage{bm}

\usepackage{mathtools}
\usepackage{physics2}
\usephysicsmodule{ab}

\usepackage{hyperref}
\hypersetup{
    colorlinks=true,
    linkcolor=blue,
    filecolor=magenta,   
    citecolor=magenta,
    urlcolor=cyan
    }

\usepackage{physics2}
\usepackage{fixdif}
\usephysicsmodule{ab}
\usepackage{siunitx}

\usepackage{color}
\usepackage{algorithm}
\usepackage{algorithmicx}
\usepackage{algpseudocode}

\makeatletter
\edef\ftype@algorithm{\the\c@float@type}
\makeatother

\begin{document}

\preprint{APS/123-QED}

\title{Adaptive Window Decoding based on Spatiotemporal Complementary Gap}

\author{Moeto Mishima}
\email{mishima.moeto@fujitsu.com}

\affiliation{
Quantum Laboratory, Fujitsu Research, Fujitsu Limited,
4-1-1 Kawasaki, Kanagawa 211-8588, Japan
}
\affiliation{
Fujitsu Quantum Computing Joint Research Division,
Center for Quantum Information and Quantum Biology, The University of Osaka, 1-2 Machikaneyama, Toyonaka, Osaka, 565-8531, Japan
}

\author{Riki Toshio}
\affiliation{
Quantum Laboratory, Fujitsu Research, Fujitsu Limited,
4-1-1 Kawasaki, Kanagawa 211-8588, Japan
}
\affiliation{
Fujitsu Quantum Computing Joint Research Division,
Center for Quantum Information and Quantum Biology, The University of Osaka, 1-2 Machikaneyama, Toyonaka, Osaka, 565-8531, Japan
}

\author{Kaito Kishi}
\affiliation{
Quantum Laboratory, Fujitsu Research, Fujitsu Limited,
4-1-1 Kawasaki, Kanagawa 211-8588, Japan
}
\affiliation{
Fujitsu Quantum Computing Joint Research Division,
Center for Quantum Information and Quantum Biology, The University of Osaka, 1-2 Machikaneyama, Toyonaka, Osaka, 565-8531, Japan
}

\author{Jun Fujisaki}
\affiliation{
Quantum Laboratory, Fujitsu Research, Fujitsu Limited,
4-1-1 Kawasaki, Kanagawa 211-8588, Japan
}
\affiliation{
Fujitsu Quantum Computing Joint Research Division,
Center for Quantum Information and Quantum Biology, The University of Osaka, 1-2 Machikaneyama, Toyonaka, Osaka, 565-8531, Japan
}

\author{Hirotaka Oshima}
\affiliation{
Quantum Laboratory, Fujitsu Research, Fujitsu Limited,
4-1-1 Kawasaki, Kanagawa 211-8588, Japan
}
\affiliation{
Fujitsu Quantum Computing Joint Research Division,
Center for Quantum Information and Quantum Biology, The University of Osaka, 1-2 Machikaneyama, Toyonaka, Osaka, 565-8531, Japan
}

\author{Shintaro Sato}
\affiliation{
Quantum Laboratory, Fujitsu Research, Fujitsu Limited,
4-1-1 Kawasaki, Kanagawa 211-8588, Japan
}
\affiliation{
Fujitsu Quantum Computing Joint Research Division,
Center for Quantum Information and Quantum Biology, The University of Osaka, 1-2 Machikaneyama, Toyonaka, Osaka, 565-8531, Japan
}

\author{Keisuke Fujii}
\affiliation{
Fujitsu Quantum Computing Joint Research Division,
Center for Quantum Information and Quantum Biology, The University of Osaka, 1-2 Machikaneyama, Toyonaka, Osaka, 565-8531, Japan
}

\affiliation{
Graduate School of Engineering Science, The University of Osaka,
1-3 Machikaneyama, Toyonaka, Osaka, 560-8531, Japan
}
\affiliation{
Graduate School of Informatics, Kyoto University, Sakyo-ku, Kyoto, 606-8501, Japan
}
\affiliation{
RIKEN Center for Quantum Computing (RQC), Wako Saitama 351-0198, Japan
}

\begin{abstract}
Real-time decoding plays a crucial role in practical fault-tolerant quantum computing. Window decoding, in which the decoding problem is divided into \emph{windows}, is a promising approach. While reducing the window size is desirable for faster decoding, each window contains a buffer region whose size must typically be at least the code distance to avoid degrading the logical error rate, which limits how much the window can shrink. In this paper, we propose an adaptive decoding scheme in which window decoding is first performed with a small buffer size and a decoding confidence (soft information) is computed; if the confidence is low, the buffer size is enlarged and decoding is redone. This approach reduces the average decoding time, since most shots are decoded with a small buffer. A central challenge in realizing this scheme is that existing forms of soft information are not directly applicable to window decoding, especially with a small buffer. We address this challenge by introducing a new form of soft information, the {\it spatiotemporal complementary gap},  specifically designed for this setting. Numerical simulations demonstrate that the proposed scheme reduces the average buffer size by approximately $\qty{40}{\percent}$ while maintaining the logical error rate. Beyond the reduction of average decoding time, the proposed method is also expected to contribute to reducing the reaction time for non-Clifford operations and relaxing the hardware requirements for real-time decoding.
\end{abstract}

\maketitle


\section{Introduction}
Fault-tolerant quantum computing (FTQC) is essential for performing large-scale quantum computations reliably, since physical qubits and quantum operations are inevitably subject to noise. Among the many quantum error-correcting codes proposed for realizing FTQC, the surface code has attracted particular attention due to its compatibility with two-dimensional qubit layouts, such as those of superconducting qubits, and its high threshold~\cite{Kitaev2003,Bravyi1998,Dennis2002,Krinner2022,Google2023suppressing,google2025quantum,Horsman2012,Litinski2019}.

A fast and accurate decoder is one of the key requirements for FTQC, especially in platforms with fast gate operations such as superconducting qubits~\cite{iOlius2024review, Battistel2023review}. Decoding is the process of inferring the underlying physical errors from the syndrome measurement outcomes, which provide indirect information about the errors. If the decoder is inaccurate, the logical error rate of the encoded qubits becomes unacceptably high, and the result of the computation can no longer be trusted. Decoding speed is equally important: in particular, non-Clifford gates cannot be executed until decoding has been completed up to the corresponding point in the circuit, so slow decoding directly slows down the quantum computation itself. If, moreover, the average decoding speed cannot keep up with the rate at which syndrome measurements are produced, the so-called backlog problem arises and the computation time grows exponentially~\cite{Terhal2015Review}. Even when the decoder is fast enough on average to avoid this issue, faster decoding remains desirable, as it reduces the idle time required before each non-Clifford gate.

One approach to accelerating decoding is window decoding. Decoding the surface code reduces to solving a matching problem on a three-dimensional cubic-lattice decoding graph, and the time required to solve this matching problem grows rapidly with the size of the graph. In window decoding, the decoding graph is divided into smaller subgraphs called windows, and matching is performed within each window to assemble a global solution. Two main variants of window decoding exist. Sliding window decoding processes windows sequentially as the window slides along the decoding graph~\cite{Dennis2002}. Parallel window decoding places multiple windows over the decoding graph and processes them in parallel~\cite{bombin2023modular,Skoric2023,Tan2023}. A number of refinements and extensions have been studied, including spatial window partitioning for lattice surgery~\cite{Lin2025}, distributed decoding across multiple FPGAs~\cite{Liyanage2025}, speculation-based prediction of inter-window dependencies to reduce reaction time~\cite{Viszlai2024}, and a method that eliminates the overhead due to window overlap~\cite{chan2026snowflake,tahaab2026stream}.

While window decoding accelerates decoding by splitting a large problem into smaller, possibly parallelizable subproblems, the windows cannot be made arbitrarily small. Each window consists of two regions: a commit region, whose local decoding result is adopted as the final error estimate, and a buffer region, which serves to improve the accuracy of the estimate within the commit region. It is known that if the buffer region is too small, the estimation accuracy degrades significantly, while a buffer region of width approximately $d$ along the decoding graph (where $d$ is the code distance) suffices to reproduce the accuracy of global decoding, in which the entire decoding graph is processed at once~\cite{bombin2023modular,Skoric2023,Tan2023,Lin2025}.

Conversely, since the buffer region must be at least approximately $d$ in width to maintain the logical error rate, this lower bound limits how small the windows can be. This can cause a serious issue in lattice surgery, where the decoding graph can extend in both spatial and temporal directions. If a $d\times d\times d$ commit region is surrounded by the decoding graph extending in all six directions, in the worst case, a buffer region must be appended in every direction, resulting in a window of size $3d\times 3d\times 3d=27d^3$. Such a large window size leads to a substantial degradation of decoding speed.

\begin{figure}
    \centering
    \includegraphics[width=\linewidth]{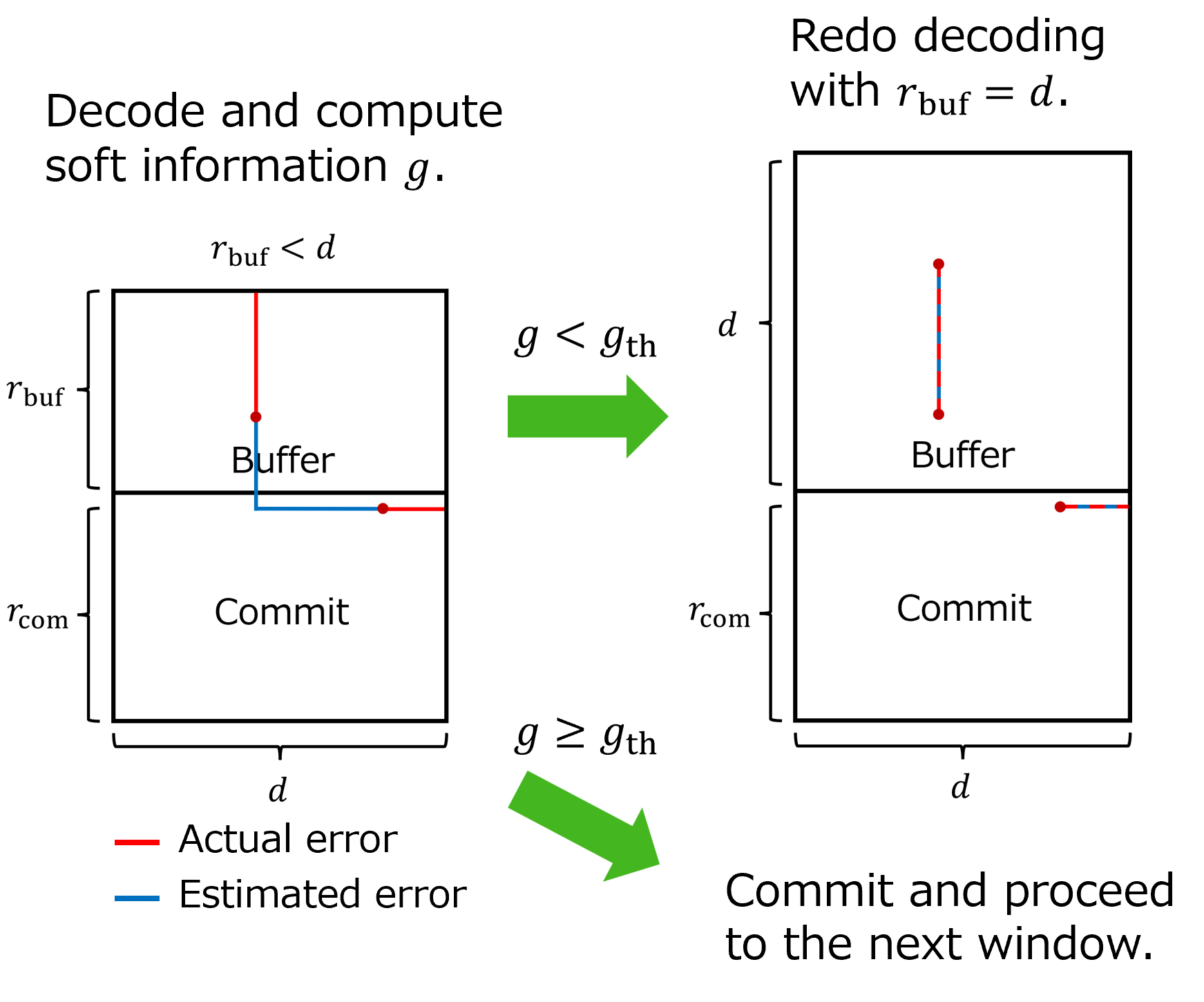}
    \caption{Schematic of the adaptive window decoding scheme. By default, decoding is performed with a small buffer size, while soft information $g$, which quantifies the decoder's confidence, is computed simultaneously. If $g$ is smaller than a predetermined threshold $g_{\mathrm{th}}$ (i.e., the decoding result is deemed unreliable), the buffer size is enlarged to $d$ and decoding is redone; otherwise, the result is committed as is. This adaptive switching reduces the average decoding time while preventing degradation of the logical error rate.}
    \label{fig__schematic_diagram}
\end{figure}

In this work, we propose a method to reduce the effective buffer size in window decoding, which we call \emph{adaptive window decoding}, as shown in Fig.~\ref{fig__schematic_diagram}. As noted above, simply shrinking the buffer region degrades the logical error rate. Our approach is therefore as follows: we first perform decoding with a small buffer size and compute a quantity that quantifies the reliability of the decoding result, often referred to as soft information; if the soft information indicates that the result is unreliable, we redo the decoding with the buffer size restored to $d$. Soft information for decoding have been studied extensively in recent years, including proposals of new soft information and methods for their computation~\cite{Hutter2014,Bombin2024,Gidney2025yoked,Meister2024,Kishi2025,Lee2025soft,xie2026simple,staples2026scalable}, and applications to performance improvements in FTQC~\cite{Gidney2024cultivation,Smith2024mitigation,dincua2025error,zhou2025error,sunami2025entanglement,Akahoshi2025timelike,toshio2025decoder}. However, the soft information relevant to our setting---that of window decoding with a small buffer size---is not naturally captured within existing frameworks. We therefore propose three new types of soft information tailored to this setting: the \emph{spatiotemporal complementary gap} (STCG), the \emph{distance-shifted STCG}, and the \emph{path-selected STCG}. Our numerical results show that the proposed soft information shares favorable properties with existing ones, and that switching based on this soft information reduces the average buffer size by approximately $\qty{40}{\percent}$ while maintaining the logical error rate. Moreover, the techniques introduced for computing the STCGs can also be used to compute the conventional complementary gap~\cite{Hutter2014,Bombin2024,Gidney2025yoked} under window decoding. Whereas this quantity has so far been formulated only under global decoding, practical FTQC requires it to be computed in real time within window decoding. Our two-stage decoding procedure fills this gap: as shown in Appendix~\ref{sec_appendix_conventional_cgap}, it reproduces the global-decoding values of the complementary gap almost exactly.

A closely related adaptive scheme was proposed independently in Ref.~\cite{oberoi2026adapt}. While both approaches adapt the decoding procedure according to the estimated reliability of intermediate results, our work is built on an explicit characterization of the failure mechanisms caused by insufficient buffer regions in window decoding. This enables us to develop new soft-information metrics specifically tailored to this regime and to use them as effective switching criteria, thereby substantially reducing the average buffer size while preserving the logical error rate.

This work proposes a method to reduce the average buffer size in window decoding while preserving the logical error rate, and demonstrates its effectiveness numerically. These results suggest that the conventional view---that the buffer size must be fixed at approximately $d$---can be relaxed in practice. Beyond the resulting reduction in average decoding time, the proposed method opens up several promising applications, such as reducing the reaction time for non-Clifford gate operations and relaxing the hardware requirements on the decoder. Together, these advances contribute to the realization of fast and accurate decoders for FTQC.  

The remainder of this paper is organized as follows. Sec.~\ref{sec_preliminary} reviews the background on decoding of the surface code. Sec.~\ref{sec_proposed_method} presents the proposed method, with particular emphasis on the definition of the new soft information. Sec.~\ref{sec_numerical_analysis} evaluates its performance through numerical simulations and then describes an extrapolation procedure to the low physical error rate regime. Sec.~\ref{sec_discussion} discusses potential applications of the proposed method. Finally, Sec.~\ref{sec_conclusion} concludes the paper and outlines directions for future work.


\section{Preliminary}
\label{sec_preliminary}

\subsection{Decoding of Surface Code}
The surface code is a quantum error-correcting code that can be implemented by arranging physical qubits on a two-dimensional square lattice. The surface code is a CSS code, whose stabilizer generators can be divided into $X$-type stabilizers, consisting only of Pauli $X$ operators, and $Z$-type stabilizers, consisting only of Pauli $Z$ operators. This structure allows the syndromes corresponding to $X$ errors and $Z$ errors to be handled separately.

Decoding is the classical process of inferring the physical errors that have occurred from measured syndrome information and determining an appropriate correction. If syndrome measurements are assumed to be ideal, the error locations can be inferred from a single round of syndrome measurements. Under a general circuit-level noise model, however, errors occur not only on data qubits but also in the syndrome measurements themselves. Consequently, the syndrome information obtained from a single measurement round cannot be fully trusted, and decoding must be performed using the syndrome history over multiple rounds.

In the following, we represent syndrome measurement outcomes as binary variables taking values in $\{0, 1\}$. A detector at $r$-th round is defined as the exclusive OR  of the syndrome measurement outcomes at the $r$-th and $(r-1)$-th  rounds corresponding to the same stabilizer. In the absence of errors, consecutive syndrome measurement outcomes agree, and hence the detector value is $0$. 
Conversely, if a physical error or a measurement error causes a change in the syndrome outcome between the two rounds, the corresponding detector takes the value $1$. In this paper, a detector whose value is $1$ is referred to as a defect.

For the standard syndrome extraction circuits of surface codes, even under circuit-level noise, elementary error events can be decomposed in such a way that each error event flips at most two detectors. This allows us to define a graph whose nodes correspond to detectors and whose edges correspond to elementary error events. This graph is called the decoding graph. An error event that flips two detectors is represented by an edge connecting the two corresponding detector nodes. For error events that flip only a single detector, we introduce a virtual node, called a boundary node, and represent such an event as an edge connecting the corresponding node to this boundary node.

Under the decomposition of error events described above, the decoding graph of a surface code can be constructed independently for the $X$-type and $Z$-type syndromes. Since the surface code is defined on a two-dimensional lattice and the multiple rounds of measurement are treated as a time direction, the decoding graph has a three-dimensional cubic lattice structure consisting of two spatial dimensions and one temporal dimension. Under the phenomenological noise model, which considers only data qubit errors and measurement errors, the resulting decoding graph forms a simple cubic lattice. Under circuit-level noise, hook errors arising during the syndrome extraction circuit give rise not only to edges along the cubic-lattice axes but also to diagonal edges in spacetime \cite{kishony2026surface, hirai2026no}. A simpler, one-dimensional analog of the surface code is the repetition code. Under the phenomenological noise model, the repetition code yields a two-dimensional square-lattice decoding graph with one spatial and one temporal dimension. Although the repetition code does not protect against both $X$ and $Z$ errors, it is often used as a simplified setting for studying the surface code.

The boundaries of the decoding graph are classified into open boundaries and closed boundaries. Nodes adjacent to an open boundary are connected to a boundary node, representing error events that flip only a single detector. In contrast, nodes on a closed boundary are not connected to any boundary node and are connected only to other ordinary nodes of the decoding graph.

Let  $E$ denote the actual error chain that has occurred and $E_{\mathrm{est}}$ denote the correction chain estimated by the decoder. The observed defects appear as the endpoints of $E$. The objective of the decoder is to estimate a correction chain  $E_{\mathrm{est}}$ whose endpoints coincide with the same defect configuration. The success of decoding is determined by the symmetric difference $E\oplus E_{\mathrm{est}}$ between the actual error chain and the correction chain. If  $E\oplus E_{\mathrm{est}}$ contains a nontrivial chain connecting inequivalent open boundaries, a logical operator is applied to the encoded state, resulting in a logical error. Otherwise, decoding succeeds.

If the probability of elementary error event $i$ is $p_i$, the corresponding edge in the decoding graph can be assigned the weight $w_i=\log((1-p_i)/p_i)$. Assuming independent error mechanisms, maximizing the likelihood of an error pattern is equivalent to minimizing the sum of the corresponding edge weights. The minimum-weight perfect matching (MWPM) algorithm~\cite{Edmonds1965_1,Wu2023fusion,Higgott2023sparse} is a representative decoding method that efficiently finds a correction chain of minimum total weight that is consistent with the observed defect configuration. Other known decoders include variants of MWPM~\cite{fowler2013optimal, delfosse2014decoding}, Union-Find~\cite{Delfosse2021UF,Liyanage2024FPGA}, which is faster but less accurate, belief-matching~\cite{Higgott2023belief_matching}, and neural-network decoders~\cite{Bausch2024alphaqubit}, which offer higher accuracy. In this paper, we focus on MWPM as a representative and analytically tractable decoder.

For the sake of the subsequent discussion, in this paper, we adopt the following convention regarding the decoding graph and the input to the MWPM. We denote the decoding graph by $G=(\mathcal{V},\mathcal{E})$, where $\mathcal{V}$ is the set of nodes and $\mathcal{E}$ is the set of edges, and each edge $e=(u,v) \in\mathcal{E}$ connects two nodes $u, v\in\mathcal{V}$. The edge weights are represented by a function $w:\mathcal{E} \to \mathbb{R}_{\geq0}$. In this paper, we treat the MWPM as a black box that takes as input the decoding graph $G$ and edge weight $w$, together with a ternary label $\sigma\in\{0,1,\bot\}^{\mathcal{V}}$ assigned to each node. Here, $\sigma(v)=0$ and $\sigma(v)=1$ indicate that the value of the detector corresponding to the node $v$ is $0$ and $1$, respectively, while $\sigma(v) = \bot$ indicates that the node $v$ is treated as a boundary node, namely that decoding is performed under the condition that the value of the detector corresponding to $v$ may be regarded as either $0$ or $1$. Under this convention, the information of whether a given node is a boundary node is encoded in the input label $\sigma$ rather than in the structure of the decoding graph $G$. Nevertheless, for notational convenience, we will sometimes refer to nodes that are assigned the label $\bot$ in the decoding procedure currently under consideration as boundary nodes.

\subsection{Window Decoding}
\begin{figure}
    \centering
    \includegraphics[width=\linewidth]{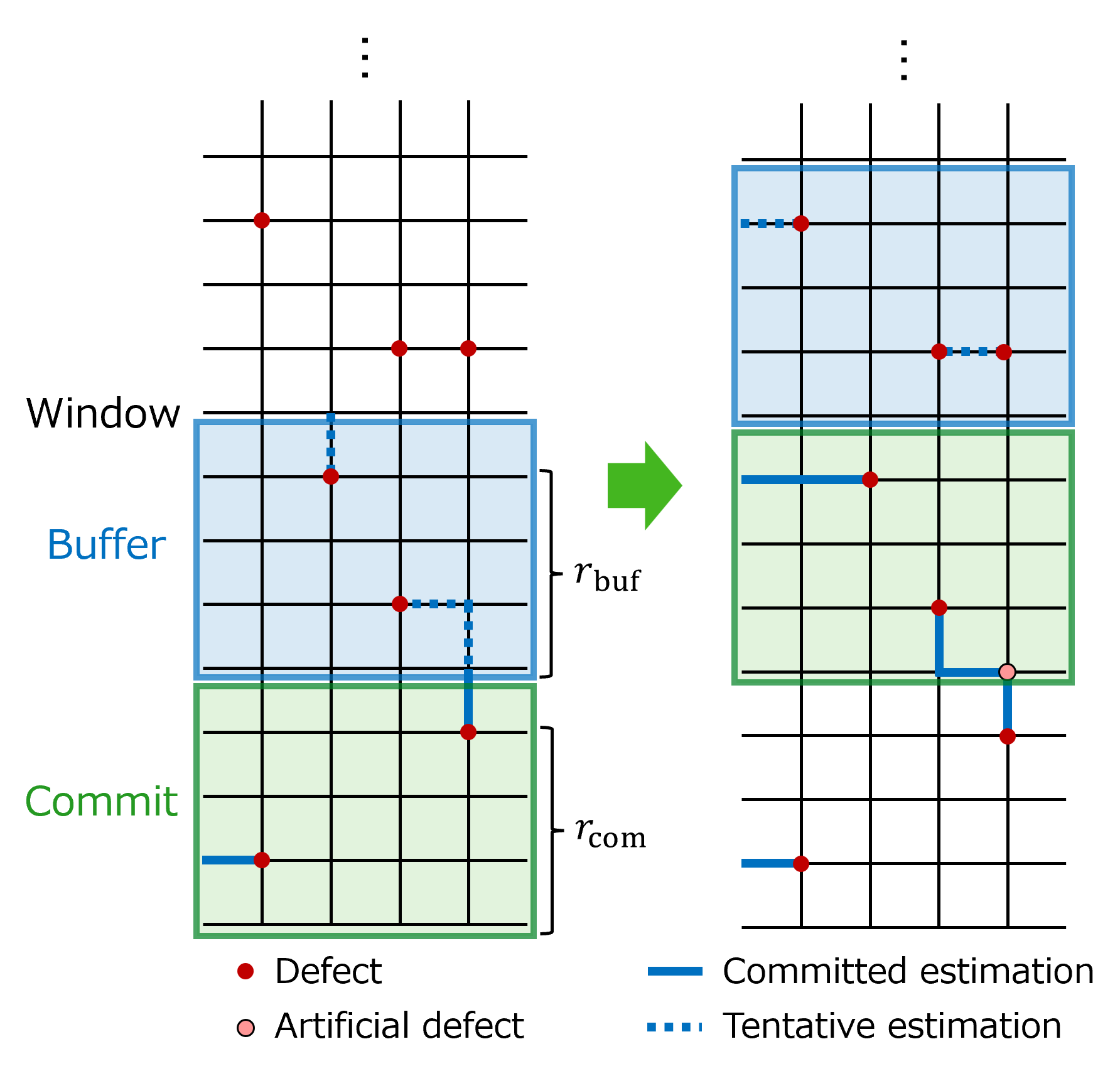}
    \caption{Schematic of sliding-window decoding. Each window consists of a commit region and a buffer region. After decoding the window, the tentative error estimate in the buffer region is discarded, and only that in the commit region is finalized. Artificial defects are generated according to the finalized error estimate in the commit region, and the window advances by $r_{\mathrm{com}}$ before the next decoding step.}
    \label{fig_sliding_window}
\end{figure}
In fault-tolerant quantum computing, decoding must be completed at least by the time a non-Clifford gate is executed. However, if one accumulates all syndrome measurement outcomes between two consecutive non-Clifford gates and decodes them in a single batch, the decoding latency becomes prohibitively large. Sliding window decoding addresses this issue by dividing the decoding graph and performing decoding each time a certain number of syndrome measurement rounds have been accumulated, thereby reducing the decoding latency. A schematic illustration of sliding window decoding is shown in Fig.~\ref{fig_sliding_window}. Each divided subgraph is referred to as a window, which consists of two regions: the commit region and the buffer region. The commit region is the region in which the estimated errors are finalized; that is, the estimated errors within the commit region constitute the output of sliding window decoding for that window. The buffer region is an additional region introduced to ensure the accuracy of the estimation in the commit region.

In what follows, we describe the sliding window decoding procedure for the surface code. For concreteness, we consider a memory experiment and present the procedure in the offline decoding setting, in which all syndrome measurement outcomes are available in advance; the procedure carries over straightforwardly to the online decoding setting, where syndrome outcomes are generated sequentially. Let $r_{\mathrm{com}}$ and $r_{\mathrm{buf}}$ denote the number of rounds in the commit region and the buffer region, respectively.

We first construct a window from the original decoding graph by extracting the nodes corresponding to the first $r_{\mathrm{com}} + r_{\mathrm{buf}}$ rounds, together with their associated defects and the edges incident to these nodes. The edges that connect nodes in the $(r_{\mathrm{com}} + r_{\mathrm{buf}})$-th round to nodes in the $(r_{\mathrm{com}} + r_{\mathrm{buf}} + 1)$-th round in the original decoding graph are reattached so that they connect the corresponding $(r_{\mathrm{com}} + r_{\mathrm{buf}})$-th round nodes to a boundary node. As a result, in addition to the spatial open boundaries inherited from the original decoding graph, the window has an additional open boundary on the future side; we refer to this temporal open boundary as the \emph{virtual boundary}. The commit region of the window consists of the nodes in the first $r_{\mathrm{com}}$ rounds and the edges incident to them, while the remaining nodes and edges constitute the buffer region. We then run MWPM on this window to obtain an estimated error. The estimated errors within the commit region are then finalized as part of the overall estimate.

We next apply the same procedure to the rounds from $r_{\mathrm{com}} + 1$ to $2 r_{\mathrm{com}} + r_{\mathrm{buf}}$ in the original decoding graph; that is, we shift the window forward by $r_{\mathrm{com}}$ rounds. There is, however, one modification: for any node in the $(r_{\mathrm{com}} + 1)$-th round that is incident to an edge in the previous window's commit region on which an error was estimated, we flip the detector value at that node. Specifically, if a defect already exists at that node, it is removed; otherwise, a new defect is placed there. We refer to defects generated in this manner as artificial defects. Decoding is then performed on the new window with this modified defect configuration in the same way as before, and the estimated errors in its commit region are finalized.

Sliding window decoding consists of repeatedly applying this procedure: shifting the window by $r_{\mathrm{com}}$ rounds, decoding within each window, finalizing the estimated errors in the commit region, and generating artificial defects for the next window. In practice, both $r_{\mathrm{com}}$ and $r_{\mathrm{buf}}$ are typically chosen to be approximately $d$ rounds. It has been shown both numerically and analytically that taking $r_{\mathrm{buf}}\simeq d$ is sufficient to achieve the same logical error rate as global decoding, in which the decoding graph is decoded at once, without being divided into windows~\cite{bombin2023modular,Skoric2023,Tan2023,Lin2025}.

\begin{figure}
    \centering
    \includegraphics[width=0.6\linewidth]{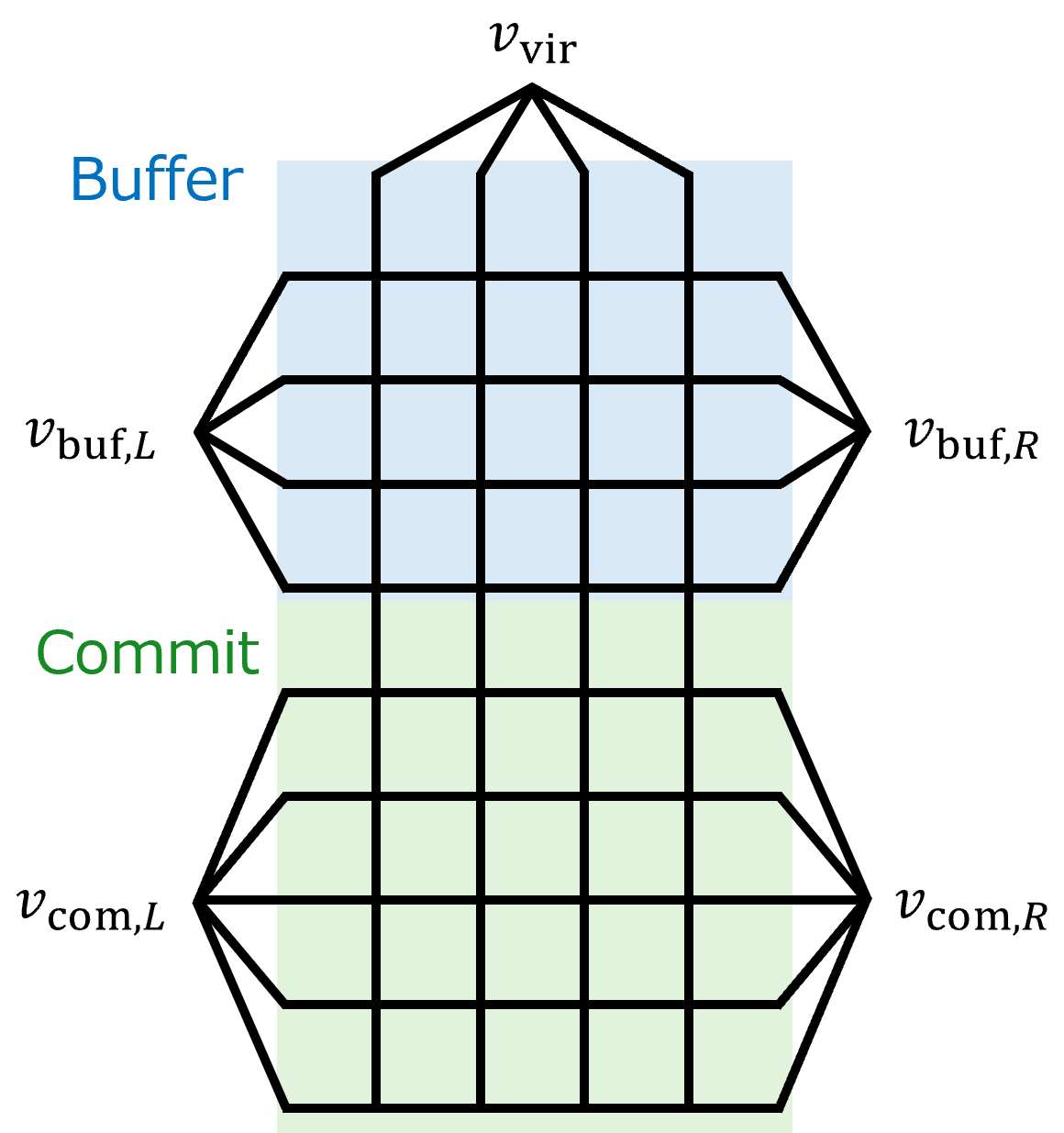}
    \caption{Schematic illustration of the boundary nodes used in the window. Separate boundary nodes are introduced for the two boundaries of the commit region ($v_{\mathrm{com},R}$ and $v_{\mathrm{com},L}$), the two boundaries of the buffer region ($v_{\mathrm{buf},R}$ and $v_{\mathrm{buf},L}$), and the virtual boundary ($v_{\mathrm{vir}}$). When performing decoding, the label $\bot$ is assigned to each of these five boundary nodes.}
    \label{fig_window}
\end{figure}

For later convenience, we assume that separate boundary nodes are introduced for the two boundaries of the commit region, for the two boundaries of the buffer region, and for the virtual boundary; see Fig.~\ref{fig_window}.

A related approach is parallel window decoding~\cite{bombin2023modular,Skoric2023,Tan2023}. Whereas sliding window decoding performs decoding sequentially by shifting a window across the decoding graph, parallel window decoding processes the decoding graph in two stages, each of which decodes multiple windows in parallel. In the first stage, multiple windows, each equipped with buffer regions, are placed across the decoding graph and decoded in parallel. The commit regions of these windows are then finalized. In the second stage, the regions remaining outside the committed parts, which form small windows between the previously committed regions, are decoded in parallel without buffer regions. As in sliding window decoding, the buffer regions used in the first stage must have a width of approximately $d$ rounds. Although we mainly focus on sliding window decoding in this paper for simplicity, the methods discussed below can be extended to parallel window decoding in a straightforward manner, at least for memory experiments.

Here, we have described window decoding assuming that MWPM is used as the decoder for each window, referred to as the inner decoder. We note that other decoders can also be employed as inner decoders.

\subsection{Complementary Gap}
\label{subsec_comp_gap}
In this subsection, we consider global decoding. By performing additional post-processing after decoding, one can obtain a quantity that estimates the reliability of the decoding outcome. Such a quantity is often referred to as soft information. Various types of soft information and methods for computing them have been proposed. Here, we focus on the complementary gap~\cite{Hutter2014,Bombin2024,Gidney2025yoked}, which has been shown to achieve particularly high performance for the surface code. In what follows, for simplicity, we assume that every edge in the decoding graph is assigned the same error rate $p$ and unit weight.

Given a syndrome measurement outcome $\sigma$, let $E_{\mathrm{min}}(\sigma)$ denote one minimum-weight error among all errors consistent with $\sigma$. Furthermore, among the errors that are consistent with $\sigma$ and belong to a logical class different from that of $E_{\mathrm{min}}(\sigma)$, let $E_{\mathrm{comp}}(\sigma)$ denote one with minimum weight. We write $w_{\mathrm{min}}(\sigma)=\mathrm{wt}(E_{\mathrm{min}}(\sigma))$ and
$w_{\mathrm{comp}}(\sigma)=\mathrm{wt}(E_{\mathrm{comp}}(\sigma))$, where $\mathrm{wt}(E)$ denotes the total weight of the edges in $E$.
The complementary gap is defined as
\begin{align}
    g(\sigma) \coloneqq w_{\mathrm{comp}}(\sigma)-w_{\mathrm{min}}(\sigma).
\end{align}

Assuming that MWPM outputs $E_{\mathrm{min}}(\sigma)$, the conditional logical error rate $P_L(\sigma)$ given the syndrome measurement outcome $\sigma$ can be expressed by considering the ratio between the probability that an error in a logical class different from that of $E_{\mathrm{min}}(\sigma)$ occurs and the probability of simply obtaining $\sigma$ as
\begin{align} \label{conditionedLER_approx_offline}
    P_L(\sigma)&\simeq\frac{n_{\mathrm{comp}}(\sigma)p^{w_{\mathrm{comp}}(\sigma)}+o(p^{w_{\mathrm{comp}}(\sigma)})}{n_{\mathrm{min}}(\sigma)p^{w_{\mathrm{min}}(\sigma)}+o(p^{w_{\mathrm{min}}(\sigma)})}\notag\\
    &\simeq\frac{n_{\mathrm{comp}}(\sigma)}{n_{\mathrm{min}}(\sigma)}p^{g(\sigma)},
\end{align}
where $n_{\mathrm{min}}(\sigma)$ is the number of minimum-weight error patterns consistent with $\sigma$, and $n_{\mathrm{comp}}(\sigma)$ is the number of minimum-weight error patterns that are consistent with $\sigma$ and belong to a logical class different from that of $E_{\mathrm{min}}(\sigma)$. We assume that $p$ is sufficiently small and use $p/(1-p)\simeq p$. If the factor
$n_{\mathrm{comp}}(\sigma)/n_{\mathrm{min}}(\sigma)$
does not vary significantly with $\sigma$, then the conditional logical error rate given $\sigma$ is larger when $g(\sigma)$ is smaller, and smaller when $g(\sigma)$ is larger. Thus, $g(\sigma)$ can be regarded as a confidence measure for the decoding result. Numerical calculations have indeed shown that the conditional logical error rate decreases exponentially with $g$~\cite{Gidney2025yoked}. This indicates that, for a given syndrome condition, the minimum-weight structure of errors alone can explain the conditional logical error rate with sufficient accuracy.

To compute $g(\sigma)$ in practice, it is enough to compute $E_{\mathrm{min}}(\sigma)$ and $E_{\mathrm{comp}}(\sigma)$. This can be done as follows. We assume here that the boundary nodes of the decoding graph are defined separately for topologically inequivalent boundaries. First, one performs the usual MWPM procedure to obtain $E_{\mathrm{min}}(\sigma)$. Suppose that, in $E_{\mathrm{min}}(\sigma)$, the boundary nodes $v_{b,0}$ and $v_{b,1}$ are matched to $n_0$ and $n_1$ defects, respectively, modulo 2. Then, for an error chain belonging to a logical class different from that of $E_{\mathrm{min}}(\sigma)$, the boundary nodes $v_{b,0}$ and $v_{b,1}$ must be matched to $n_0\oplus 1$ and $n_1\oplus 1$ defects, respectively, modulo 2. Therefore, by rerunning MWPM with
$\sigma(v_{b,0})=n_0\oplus 1$ and
$\sigma(v_{b,1})=n_1\oplus 1$
while keeping all other defect configurations unchanged, one obtains the minimum-weight error $E_{\mathrm{comp}}(\sigma)$ in the different logical class.

\section{Adaptive Window Decoding}
\label{sec_proposed_method}
\subsection{Key Issue}
The method we propose in this paper for reducing the size of the decoding problem is as follows (see also Fig.~\ref{fig__schematic_diagram}): in window decoding, we first set $r_{\mathrm{buf}}$ to a value smaller than $d$ and compute the soft information $g$. If $g$ is smaller than a predetermined threshold, that is, if the decoding result is deemed unreliable, we redo the decoding with $r_{\mathrm{buf}}=d$. A key issue in this approach is how to define the soft information in window decoding.

\begin{figure*}
    \centering
    \includegraphics[width=0.8\linewidth]{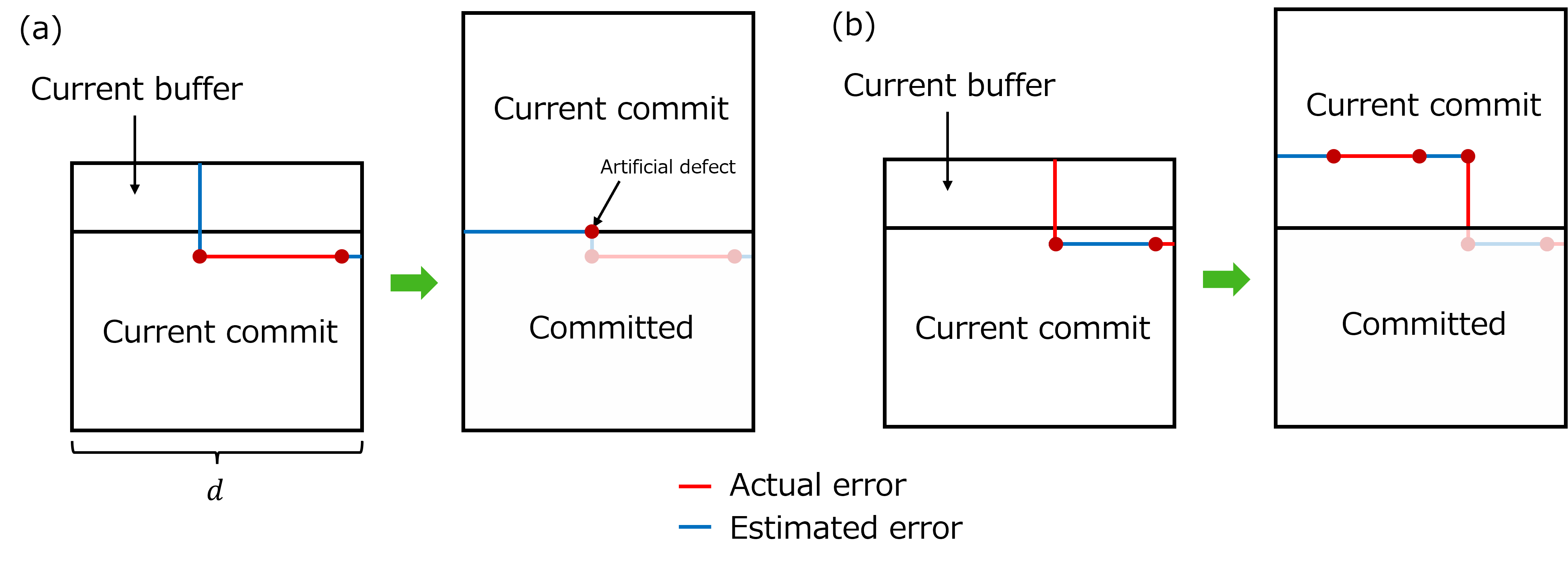}
    \caption{Examples of logical errors caused by a small buffer size. Red lines represent the actual error chains, and blue lines represent the estimated error chains. (a) Two defects in window $i$ are incorrectly matched to the right commit boundary and the virtual boundary, leaving an artificial defect with no partner in window $i+1$, which is then matched to the left commit boundary by MWPM. (b) Two defects that should have been matched to the boundaries are instead matched to each other in window $i$, leading to an unanticipated single defect in window $i+1$, which is then matched to the left boundary through a pair of defects.}
    \label{fig_logical_error_example}
\end{figure*}

In global decoding, the close relationship between the complementary gap and the conditional logical error rate is relatively straightforward, as shown in Sec.~\ref{subsec_comp_gap}. In window decoding, however, the situation is more involved. Consider, for example, a case in which the error estimation in window $i$ is incorrect, yet no logical error occurs when the estimated and actual errors are combined. Even in this case, a logical error may still arise after decoding window $i+1$, through the combination of estimated and actual errors across windows $i$ and $i+1$, as a consequence of the incorrect estimation in window $i$ (see also Fig.~\ref{fig_logical_error_example}). Such mechanisms of logical error in window decoding cannot be captured simply by re-running MWPM with modified boundary conditions, as in the computation of the complementary gap. A more sophisticated method is therefore required to approximate the conditional logical error rate.

Let us first consider an example in which sliding window decoding with a small buffer size fails to produce the minimum-weight error. In Fig.~\ref{fig_logical_error_example} (a), a logical error occurs in window $i+1$ as a consequence of incorrect estimation in window $i$. In window $i$, two defects are incorrectly matched to the right commit boundary and to the virtual boundary, leaving an artificial defect with no partner in window $i+1$. When MWPM is performed in this situation, the artificial defect is matched to the left commit boundary, resulting in a logical error. Figure~\ref{fig_logical_error_example} (b) shows another example, in which two defects that should have been matched to the boundaries are instead matched to each other. Consequently, in window $i+1$, a single defect appears that was not anticipated by the estimation in window $i$. Under MWPM, this defect is effectively matched to the left boundary through a pair of defects produced by an error chain, again resulting in a logical error.

In both examples, logical errors arise in sliding window decoding through the following mechanism. An incorrect treatment of the virtual boundary in window $i$---either matching a defect to the virtual boundary when it should not be matched there, or failing to match a defect that should be matched there---leads to an incorrectly committed error. As a consequence, a defect is left without a partner in window $i+1$, where it is then matched to an incorrect boundary.

\subsection{Soft Information in Window Decoding}
\subsubsection{General Theory} \label{subsubsec_general_theory}
In this section, we introduce quantities that serve as indicators of the conditioned logical error rate, given a syndrome measurement outcome within a window under sliding window decoding. The underlying idea is analogous to that of the complementary gap. As discussed in Sec.~\ref{subsec_comp_gap}, the complementary gap provides a reliable estimate of the conditioned logical error rate under the global decoding by considering only the minimum-weight error patterns for different logical classes. We therefore adopt a similar approach here, focusing on the minimum weight while neglecting the degeneracies and higher-order terms.

We are concerned here with logical errors that arise specifically from the use of a small buffer size. Accordingly, we do not consider logical errors that are confined to the commit region of the current window or to the regions preceding it. We further assume that the decoding in the previous window is consistent with the global decoding; that is, the artificial defects are generated correctly. We adopt a phenomenological noise model on the repetition code or the surface code, and assume that all edges in the decoding graph are assigned a uniform error rate $p$ and unit weight. An extension to a circuit-level noise model on the surface code is briefly discussed later.

Let $\sigma$ denote the syndrome measurement outcome obtained in a given window. The probability $P(\sigma)$ that $\sigma$ occurs can be approximated, in terms of the weight $w_{\mathrm{min}}$ of the MWPM solution $E_{\mathrm{min}}$ in the window, as
\begin{align}
P(\sigma) \sim p^{w_{\mathrm{min}}},
\end{align}
where the symbol $\sim$ indicates that the degeneracy of error patterns and the higher-order terms are neglected assuming that $p$ is sufficiently small.

Next, we evaluate the joint probability $P(\sigma \cap \mathrm{fail})$ that the syndrome measurement outcome $\sigma$ is observed and a logical error occurs in the subsequent window. Among all error patterns consistent with $\sigma$, we seek the one that contributes most to the logical error rate. In the case of the complementary gap, the analogous role is played by the minimum-weight error pattern in the opposing logical class of $E_{\mathrm{min}}$, which admits a simple conceptual characterization. In the present setting, however, no such simple characterization is available. We therefore proceed by considering an arbitrary error pattern consistent with $\sigma$ and distinct from $E_{\mathrm{min}}$, which we denote $E_{\mathrm{alt}}$ with weight $w_{\mathrm{alt}}$, and suppose that $E_{\mathrm{alt}}$ is the actual error in the current window. Let $t_{\mathrm{min}}$ denote the minimum weight of an error in the subsequent window required for a logical error to occur, given that $E_{\mathrm{min}}$ has been committed and MWPM is then performed in the subsequent window. Note that $t_{\mathrm{min}}$ depends on $E_{\mathrm{alt}}$. With these definitions, the probability that the actual error is $E_{\mathrm{alt}}$ and that a logical error occurs can be expressed as $\sim p^{w_{\mathrm{alt}}+t_{\mathrm{min}}}$. Defining
\begin{align}
    w_{\mathrm{comp}} \coloneqq \min_{E_{\mathrm{alt}}}(w_{\mathrm{alt}}+t_{\mathrm{min}}),
\end{align}
we obtain
\begin{align}
    P_L(\sigma)&=\frac{P(\sigma \cap \mathrm{fail})}{P(\sigma)} \\ &\sim\frac{p^{w_\mathrm{comp}}}{p^{w_{\mathrm{min}}}}=p^{\min_{E_{\mathrm{alt}}} (w_{\mathrm{alt}}+t_{\mathrm{min}}-w_\mathrm{min})}.
\end{align}

Consequently, by identifying $E_{\mathrm{alt}}$ minimizing $w_{\mathrm{alt}}+t_{\mathrm{min}}$ and computing $w_{\mathrm{alt}}+t_{\mathrm{min}}-w_{\mathrm{min}}$, this quantity can serve as a soft information. In general, however, neither identifying the optimal $E_{\mathrm{alt}}$ nor evaluating $t_{\mathrm{min}}$ for a given $E_{\mathrm{alt}}$ is straightforward. We therefore propose, in the following, several algorithms for searching for $E_{\mathrm{alt}}$ that yield small values of $w_{\mathrm{alt}}+t_{\mathrm{min}}$.

\subsubsection{Spatiotemporal Complementary Gap} \label{subsubsec_STCG}
In this subsection, we introduce a concrete form of soft information tailored to window decoding with a small buffer, which we call the spatiotemporal complementary gap (STCG). As discussed in Sec.~\ref{subsubsec_general_theory}, the principal mechanism by which a logical error arises is that defects are incorrectly matched to the virtual boundary, leading to the commitment of incorrect errors (see Fig.~\ref{fig_logical_error_example}). This observation motivates a natural choice of $E_{\mathrm{alt}}$; namely, the error obtained by taking the complementary configuration with respect to the virtual boundary and the commit boundary, relative to $E_{\mathrm{min}}$ (hereafter, we denote this specific choice by $E_{\mathrm{alt,STCG}}$, with weight $w_\mathrm{alt,STCG}$, and refer to it as the complementary error). We further neglect $t_{\mathrm{min}}$, which is generally difficult to evaluate. One might object that ignoring $t_{\mathrm{min}}$ is problematic; however, as shown later, the simple quantity $g = w_{\mathrm{alt,STCG}} - w_{\mathrm{min}}$ already captures the conditioned logical error rate well, even without $t_{\mathrm{min}}$.

\begin{figure}
    \centering
    \includegraphics[width=\linewidth]{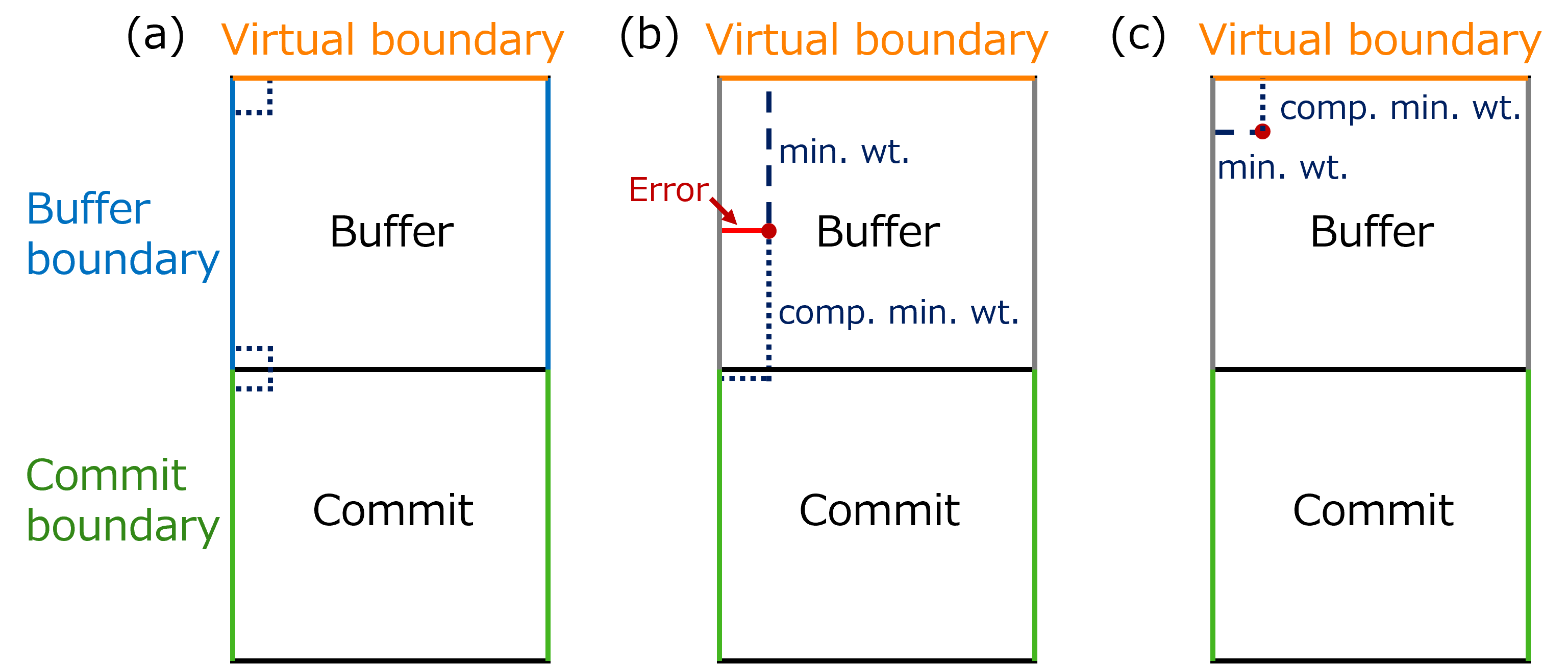}
    \caption{Difficulties in computing the complementary error within a window. (a) Leaving the buffer boundary open only yields additional errors indicated by the blue dotted lines. (b) Closing the buffer boundary for both the minimum-weight and complementary errors leads to a spuriously small gap in some cases. (c) Computing the minimum-weight error with the buffer boundary open and the complementary error with it closed also leads to a spuriously small gap in some cases.}
    \label{fig_gap_computation_problem}
\end{figure}

Computing $E_{\mathrm{alt,STCG}}$ is itself non-trivial, even after this simplification, as shown in Fig.~\ref{fig_gap_computation_problem}. A naive approach would be to rerun MWPM with the boundary conditions on the virtual and commit boundaries flipped relative to those used for the minimum-weight error. This approach, however, only yields the additional low-weight error indicated by the blue dotted line in Fig.~\ref{fig_gap_computation_problem} (a), because the virtual and commit boundaries are connected through the buffer boundary---in other words, the effective distance between them is fixed at a small value that does not depend on $d$ and $r_{\mathrm{buf}}$.

One might attempt to circumvent this issue by closing the buffer boundary. This remedy introduces a different problem, however, when a short error chain extends from the buffer boundary. If both the minimum-weight error and the complementary error are computed with the buffer boundary closed, the gap becomes spuriously small in cases such as the one illustrated in Fig.~\ref{fig_gap_computation_problem} (b). If, instead, the minimum-weight error is computed in the original window while only the complementary error is computed with the buffer boundary closed, the gap likewise becomes spuriously small in cases such as the one shown in Fig.~\ref{fig_gap_computation_problem} (c).

\begin{figure}
    \centering
    \includegraphics[width=\linewidth]{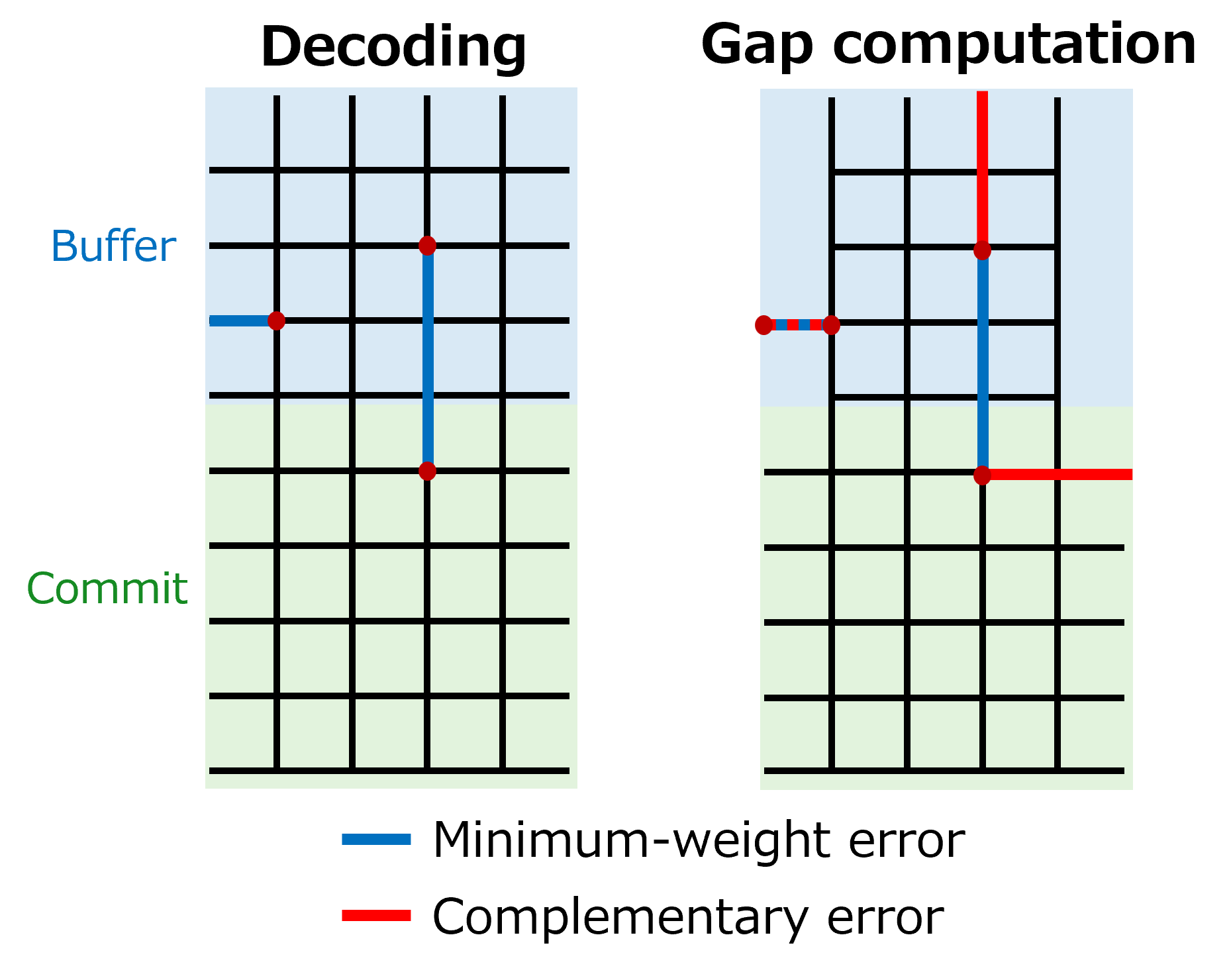}
    \caption{Procedure for computing the STCG. After the minimum-weight error is first obtained, the buffer boundary is modified, and the gap is then computed.}
    \label{fig_STCG_computation}
\end{figure}

We therefore propose the following procedure for computing $E_{\mathrm{alt,STCG}}$, as shown in Fig.~\ref{fig_STCG_computation}. First, the minimum-weight error is computed in the original window. Next, the buffer boundary is closed; that is, the edges connecting bulk nodes to the boundary node along the buffer boundary are removed. As an exception, if the minimum-weight error includes edges on the buffer boundary---equivalently, if defects are matched to the buffer boundary in the minimum-weight error---those edges are retained, with the boundary node replaced by a new node at which an additional defect is placed. Retaining these edges is a key difference from the procedure shown in Fig.~\ref{fig_gap_computation_problem}~(c). The complementary minimum-weight error is then computed in this modified window, and the STCG is defined as the resulting gap $g \coloneqq w_\mathrm{alt,STCG}-w_\mathrm{min}$.

Even if a defect matched to the buffer boundary should in fact have been matched to another defect or to a different boundary, such an incorrect matching rarely causes a logical error. The STCG therefore effectively disregards error chains connected to the buffer boundary when the complementary error is computed. We note that, strictly speaking, a defect matched to the buffer boundary in the minimum-weight error is not necessarily matched to the corresponding additional defect in the complementary error. We also note that the minimum-weight error in the modified window coincides with that in the original window.

\emph{Remark:} A similar difficulty arises when one attempts to compute the conventional complementary gap (Sec.~\ref{subsec_comp_gap}) in window decoding. In this case, the virtual boundary connects the left and right spatial boundaries, so that their effective distance is reduced to a small value independent of $d$, whereas naively closing the virtual boundary yields spuriously small gaps for reasons analogous to those illustrated in Fig.~\ref{fig_gap_computation_problem}. The modification of the window introduced above---closing a boundary irrelevant to the gap under consideration while retaining the edges used by $E_\mathrm{min}$ and placing additional defects---can be regarded as a general technique for eliminating the effect of such an irrelevant boundary on gap computation. Applying it to the virtual boundary enables the conventional complementary gap to be computed in window decoding; see Appendix \ref{sec_appendix_conventional_cgap} for details.

\subsubsection{Distance-shifted STCG} \label{subsubsec_dsSTCG}
In Sec.~\ref{subsubsec_STCG}, we considered the quantity $w_{\mathrm{alt,STCG}}-w_{\mathrm{min}}$ while neglecting $t_{\mathrm{min}}$. As noted there, computing the exact value of $t_{\mathrm{min}}$ for general error configuration $E_{\mathrm{min}}$ and $E_{\mathrm{alt}}$ is difficult. We therefore introduce a simple approximation for $t_{\mathrm{min}}$.

\begin{figure}
    \centering
    \includegraphics[width=\linewidth]{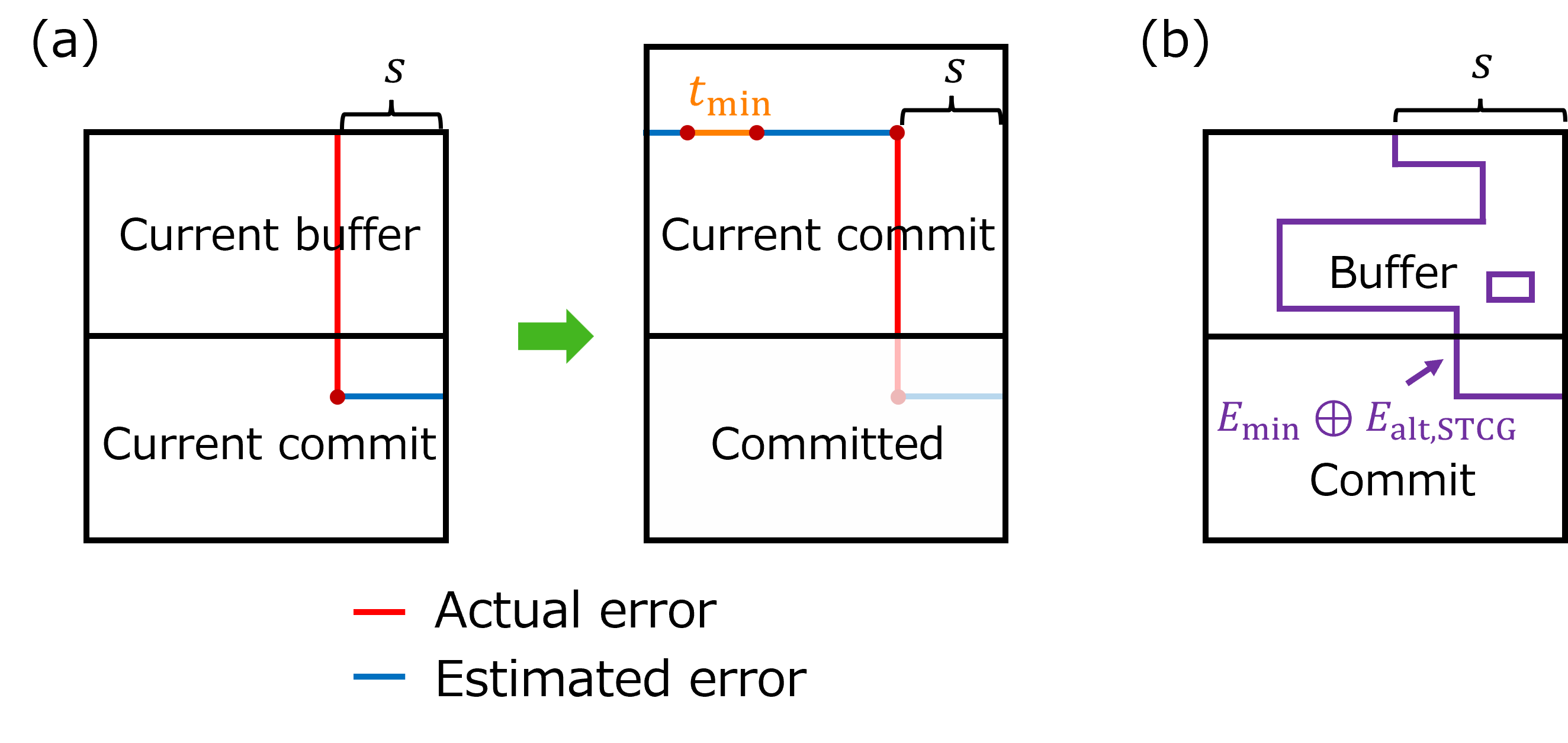}
    \caption{(a) $t_{\mathrm{min}}$ in a simple case. (b) The distance $s$ used in computing the distance-shifted STCG.}
    \label{fig_illustration_s_and_t_min}
\end{figure}

We first consider a simple example in which only a single defect is present in the window, as illustrated in Fig.~\ref{fig_illustration_s_and_t_min} (a). Suppose that, in the minimum-weight error, this defect is matched to the right boundary with weight $s$, while in reality an error chain connecting the defect to the virtual boundary has occurred. In this case, $t_{\mathrm{min}}$ can be evaluated explicitly. The minimum-weight scenario is one in which the error chain appears at the round right after the committed window. The condition for $t_{\mathrm{min}}$ to lead to a logical error is that a single defect generated by the incorrect estimation is (effectively) matched to the left boundary when MWPM is performed in the next window, namely, $(d-s)-t_{\mathrm{min}} \leq s + t_{\mathrm{min}}$. This yields $t_{\mathrm{min}}=\max(\lceil d/2\rceil -s,0)$.

We now propose adopting this expression, $t_{\mathrm{min}}=\max(\lceil d/2\rceil -s,0)$ as an approximation in the general case. First, we compute $E_{\mathrm{min}}$ and $E_{\mathrm{alt,STCG}}$ following the procedure for the STCG. Then, the error chain $E_{\mathrm{min}}\oplus E_{\mathrm{alt,STCG}}$ consists of a path connecting the virtual and commit boundaries together with trivial cycles. Suppose that this path terminates at the right commit boundary, and let $s$ denote the distance between the right boundary and the point at which the path meets the virtual boundary. We then define 
\begin{align}
    g \coloneqq w_{\mathrm{alt,STCG}} + \max (\lceil d/2 \rceil - s, 0) - w_{\mathrm{min}}
\end{align}
Even in the general case, the probability of a logical error occurring in the next window is expected to increase with $s$, and the proposed approximation reflects this intuition. We refer to this quantity as the \emph{distance-shifted STCG}.

\subsubsection{Path-selected STCG} \label{subsubsec_psSTCG}
In computing distance-shifted STCG, we first determine $E_{\mathrm{alt,STCG}}$ as the complementary error between the virtual and commit boundaries with respect to $E_{\mathrm{min}}$, and then compute $t_{\mathrm{min}}$ approximately. While this procedure yields an $E_{\mathrm{alt,STCG}}$ of small weight, the corresponding $t_{\mathrm{min}}$ may not be small. In other words, there may exist another choice of $E_{\mathrm{alt}}$ with a slightly larger weight but a smaller $t_{\mathrm{min}}$ leading to a smaller value of $w_{\mathrm{alt}} + t_{\mathrm{min}}$. To tackle this issue, we propose a method that minimizes $w_{\mathrm{alt}} + t_{\mathrm{min}}$ directly, with $t_{\mathrm{min}}$ approximated by $\max(\lceil d/2\rceil -s,0)$ as in the previous subsection. The minimization itself is also performed approximately, as detailed below.

\begin{figure}
    \centering
    \includegraphics[width=\linewidth]{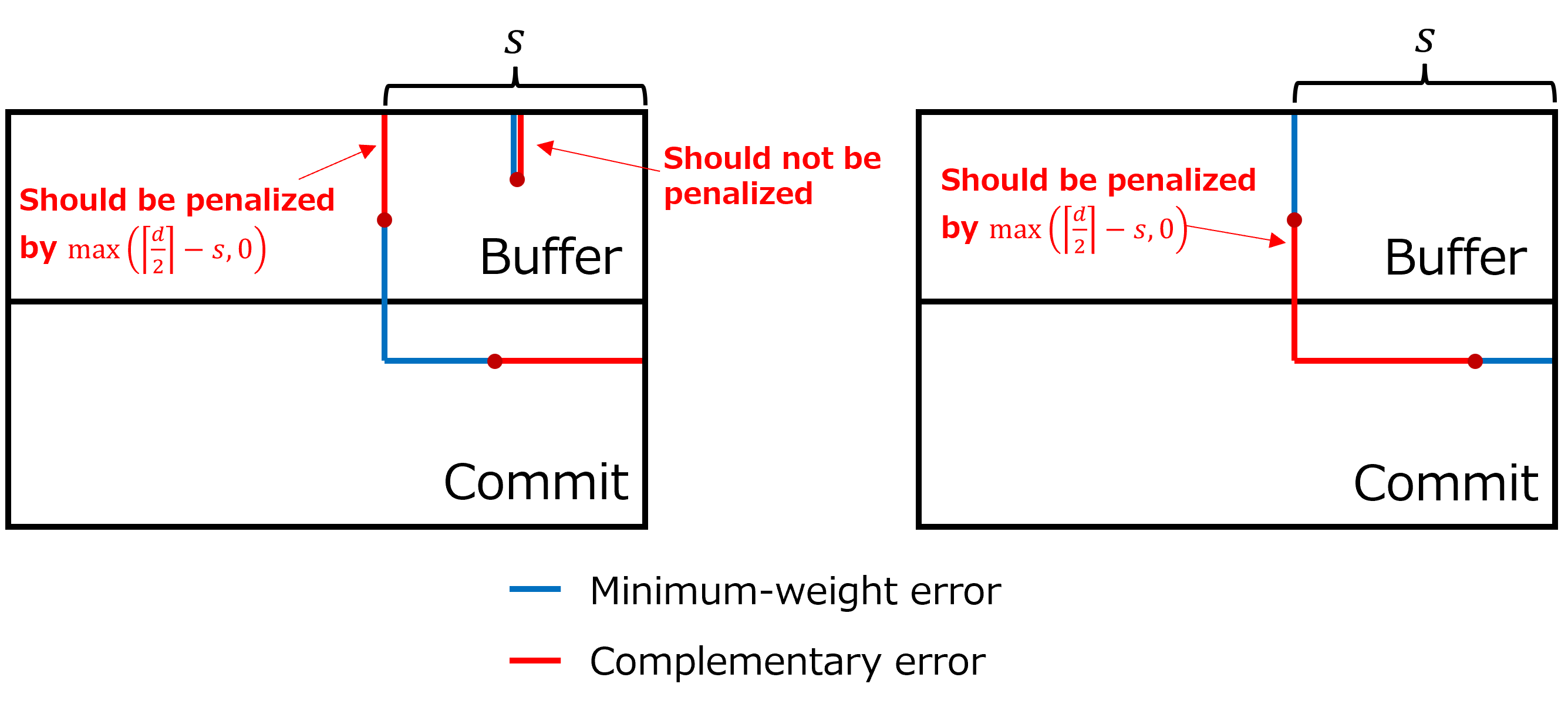}
    \caption{Error chains that should be penalized and those that should not.}
    \label{fig_which_error_chain_to_be_penalized}
\end{figure}

The minimization is implemented by modifying the decoding graph used to compute $E_{\mathrm{alt,STCG}}$. The basic idea is to add a penalty term corresponding to $t_{\mathrm{min}} = \max (\lceil d/2 \rceil -s,0)$ to edges in the window, although some subtleties arise in the details. As illustrated in Fig.~\ref{fig_which_error_chain_to_be_penalized}, given the minimum weight error in the window, the penalty $\max (\lceil d/2 \rceil -s, 0)$ should be added in two cases: (i) when a defect that was matched to another defect in $E_{\mathrm{min}}$ is matched to the virtual boundary in $E_{\mathrm{alt}}$, and (ii) conversely when a defect that was matched to the virtual boundary in $E_{\mathrm{min}}$ is matched to another defect in $E_{\mathrm{alt}}$. The penalty should not, however, be added (iii) when a defect matched to the virtual boundary in $E_{\mathrm{min}}$ is again matched to the virtual boundary in $E_{\mathrm{alt}}$.

\begin{figure*}
    \centering
    \includegraphics[width=0.8\linewidth]{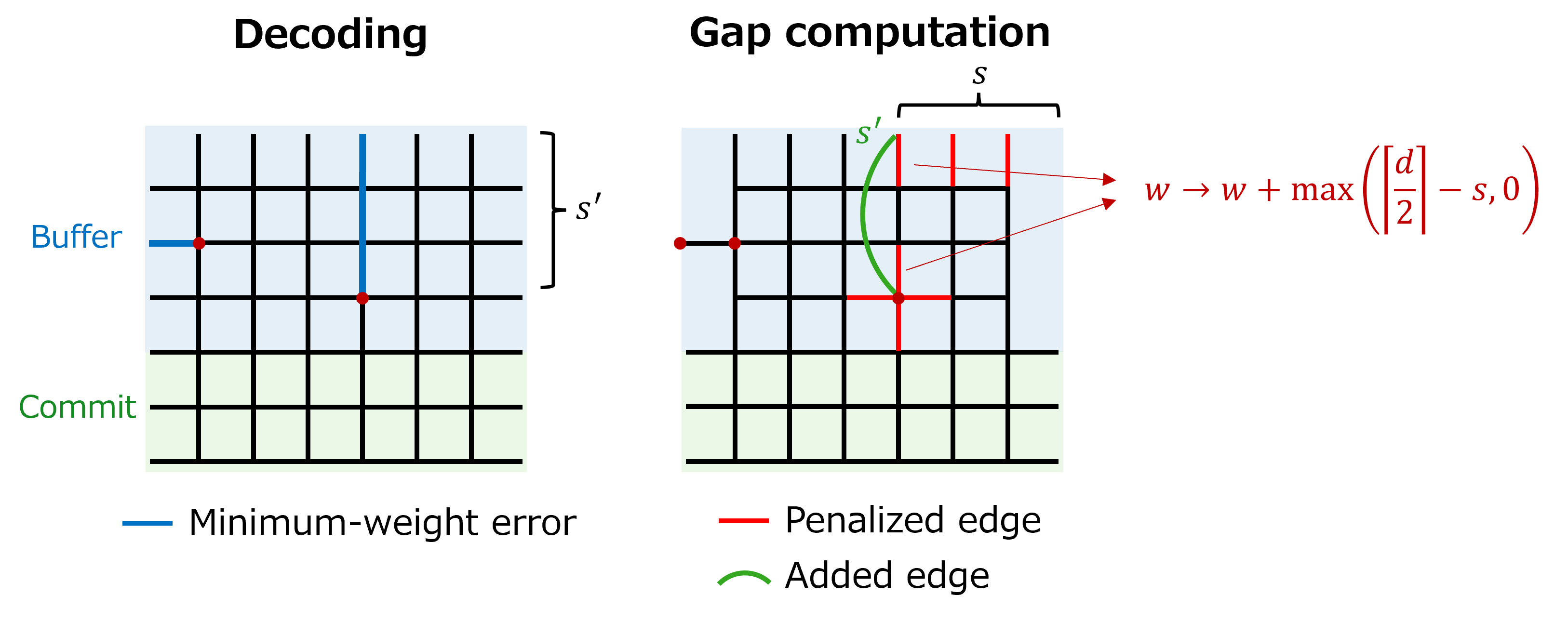}
    \caption{Procedure for computing the path-selected STCG. After the minimum-weight error is first obtained, the buffer boundary is modified, edges are added, penalties are added to edge weights, and the gap is then computed.}
    \label{fig_psSTCG_computation}
\end{figure*}

Based on these observations, we now describe how to modify the decoding graph (see also Fig.~\ref{fig_psSTCG_computation}). We first present the procedure for computing the gap from the right boundary; the corresponding procedure for the left boundary is entirely analogous. 

After computing $E_{\mathrm{min}}$ in the original window, we apply the same modification on the buffer boundary as in the STCG. We then introduce three further modifications. First, corresponding to case (i), to each edge on the virtual boundary we add a penalty $\max(\lceil d/2 \rceil- s, 0)$ to its weight, where $s$ is the distance from the right boundary to that edge. This penalizes paths in $E_{\mathrm{alt}}$ that reach the virtual boundary. Second, corresponding to case (ii), for each defect $u$ matched to the virtual boundary in $E_{\mathrm{min}}$, we add the penalty $\max(\lceil d/2 \rceil- s, 0)$ to the weight of every edge incident to $u$, where $s$ is the distance from the right boundary to $u$. This penalizes paths in $E_{\mathrm{alt}}$ in which $u$ is matched to another defect rather than to the virtual boundary, since such paths necessarily start from an edge incident to $u$. Third, corresponding to case (iii), for each such defect $u$, we add a direct edge connecting $u$ to the virtual boundary directly with weight $s'$, where $s'$ is the distance from $u$ to the virtual boundary in the original window. This edge allows $u$ to remain matched to the virtual boundary in $E_{\mathrm{alt}}$ without being subject to the penalty introduced in the second modification.

We then compute the complementary minimum-weight error with respect to $E_{\mathrm{min}}$ in this modified window. The resulting error chain provides $E_{\mathrm{alt}}$ that (approximately) minimizes $w_{\mathrm{alt}} + \max (\lceil d/2 \rceil -s, 0)$ with respect to the right boundary, and we denote its weight in this modified graph by $w'_{\mathrm{alt,R}}$. Repeating the analogous procedure for the left boundary yields $w'_{\mathrm{alt,L}}$. Taking the minimum of the two and subtracting $w_{\mathrm{min}}$, we define the gap as
\begin{align}
    g \coloneqq \min(w'_{\mathrm{alt,R}},\, w'_{\mathrm{alt,L}}) - w_{\mathrm{min}}.
\end{align}
We refer to this quantity as the \emph{path-selected STCG}.

We note again that this minimization is approximate. An exact minimization could be obtained by introducing further modifications to the decoding graph, but we expect that such refinements would offer little improvement in the gap's effectiveness as a switching criterion.

\subsubsection{Extension to Circuit-level Noise}
The discussion so far has assumed a phenomenological noise model with uniform edge weights. For practical relevance, it is important to extend the soft-output schemes to the circuit-level noise model on the surface code. The main differences from the phenomenological case are twofold: diagonal edges appear in addition to the square or cubic edges, and the edges are assigned non-uniform error rates and correspondingly non-uniform weights. Both modifications affect the definitions and computations of the STCGs. Most notably, the distances entering the penalty terms must be evaluated as shortest-path distances on the global decoding graph rather than simple lattice distances. Below, we describe how each gap is modified. Note that under the circuit-level noise model, two or more edges may exist between the virtual boundary and a single node due to the presence of diagonal edges. In such cases, we keep only the edge with the smallest weight.

The STCG is the simplest of the three to extend, as its computation involves only standard MWPM and the elimination of buffer edges. Both operations carry over directly to the circuit-level setting.

For the distance-shifted STCG, the evaluation of $t_{\mathrm{min}}$ must be modified. Recall the situation in Fig.~\ref{fig_illustration_s_and_t_min} (a), in which a single defect $u$ remains as a result of an incorrect estimation. For a logical error to occur when MWPM is performed in the next window, $t_{\mathrm{min}}$ must satisfy
\begin{align}
    d_{\mathrm{left}}(u) - t_{\mathrm{min}} \leq d_{\mathrm{right}}(u) + t_{\mathrm{min}},
\end{align}
where $d_{\mathrm{left}}(u)$ and $d_{\mathrm{right}}(u)$
denote the distances from $u$ to the left and right boundaries, respectively, in the global decoding graph. Solving for $t_{\mathrm{min}}$ yields
\begin{align}
    t_{\mathrm{min}} = \max\left(\frac{d_{\mathrm{left}}(u) - d_{\mathrm{right}}(u)}{2},\, 0\right),
\end{align}
which generalizes the phenomenological expression $t_{\mathrm{min}} = \max(\lceil d/2 \rceil - s,\, 0)$. The case in which the path $E_{\mathrm{min}}\oplus E_{\mathrm{alt,STCG}}$ terminates at the left commit boundary is treated analogously, with the roles of $d_{\mathrm{left}}$ and $d_{\mathrm{right}}$ exchanged.

For the path-selected STCG, the graph modifications introduced in Sec.~\ref{subsubsec_psSTCG} extend straightforwardly: each occurrence of $\max(\lceil d/2 \rceil - s,\, 0)$ is replaced by $t_{\mathrm{min}}$ as defined above, with the argument $u$ taken to be the relevant point in each modification. For the penalty added to edges on the virtual boundary, $u$ is taken as the endpoint of the edge; for the penalty added to edges incident to a defect matched to the virtual boundary in $E_{\mathrm{min}}$, $u$ is the defect itself. The weight $s'$ of the direct edge connecting such a defect to the virtual boundary is similarly replaced by the shortest-path distance from the defect to the virtual boundary in the original window.

The complete algorithms for computing the STCG, distance-shifted STCG, and path-selected STCG are presented in Algorithms~\ref{alg_STCG},~\ref{alg_dsSTCG}, and~\ref{alg_psSTCG}, respectively. Here, Algorithm~\ref{alg_psSTCG} computes the path-selected STCG from the right boundary. The gap from the left boundary can be computed analogously. In the algorithm, $N(\cdot)$ denotes the neighborhood for a node in the graph.

\begin{algorithm}
    \caption{Spatiotemporal complementary gap}
    \label{alg_STCG}
    \begin{algorithmic}[1]
        \Require A window graph $G=(\mathcal{V},\mathcal{E})$ with edge weight $w$, commit boundary nodes $v_{\mathrm{com},R}$ and $v_{\mathrm{com},L}$, buffer boundary nodes $v_{\mathrm{buf},R}$ and $v_{\mathrm{buf},L}$, and virtual boundary node $v_{\mathrm{vir}}$ and syndrome measurement outcome $\sigma$.
        \Ensure STCG
        \State {Compute the minimum weight error $E_{\mathrm{min}}$ and its weight $w_{\mathrm{min}}$ in $G$ with input $\sigma$.}
        \For {Each $v\in N(v_{\mathrm{buf},R})\cup N(v_{\mathrm{buf},L})$}
            \If {$v\in N(v_{\mathrm{buf},R})$}
                \State {$e\gets(v,v_{\mathrm{buf},R})$}
            \Else
                \State {$e\gets(v,v_{\mathrm{buf},L})$}
            \EndIf
            \If {$ E_{\mathrm{min}}(e)=1$}
                \State {Add a new node $u_v$ to $\mathcal{V}$ and a new edge $(v,u_v)$ to $\mathcal{E}$ with weight $w(e)$.}
                \State {$\sigma(u_v)\gets 1$}
            \EndIf
            \State {Eliminate $e$ from $\mathcal{E}$.}
        \EndFor
        \State {$\sigma(v_{\mathrm{vir}})\gets 1\oplus\bigoplus_{v\in N(v_{\mathrm{vir}})}E_{\mathrm{min}}((v,v_{\mathrm{vir}}))$}
        \State Compute the minimum weight error $E_{\mathrm{alt,STCG}}$ and its weight $w_{\mathrm{alt,STCG}}$ in $G$ with input $\sigma$.
        \State \Return $w_{\mathrm{alt,STCG}}-w_{\mathrm{min}}$
    \end{algorithmic}
\end{algorithm}

\begin{algorithm}
    \caption{Distance-shifted STCG}
    \label{alg_dsSTCG}
    \begin{algorithmic}[1]
        \Require A window graph $G=(\mathcal{V},\mathcal{E})$ with edge weight $w$, commit boundary nodes $v_{\mathrm{com},R}$ and $v_{\mathrm{com},L}$, buffer boundary nodes $v_{\mathrm{buf},R}$ and $v_{\mathrm{buf},L}$, and virtual boundary node $v_{\mathrm{vir}}$, distance $d_{\mathrm{left}}$ and $d_{\mathrm{right}}$ from each node to the left and right boundaries in the global decoding graph, and syndrome measurement outcome $\sigma$.
        \Ensure Distance-shifted STCG
        \State {Compute STCG $g_{\mathrm{STCG}}$ using Algorithm \ref{alg_STCG} recording the minimum weight error $E_{\mathrm{min}}$ and the complementary error $E_{\mathrm{alt,STCG}}$.}
        \State Compute two endpoint of the path $E_{\mathrm{min}}\oplus E_{\mathrm{alt,STCG}}$, as $u_0$ on the commit boundary and $v_0$ on the virtual boundary in the global decoding graph.
        \If {$u_0$ is on the right commit boundary}
            \State \Return $g_{\mathrm{STCG}}+\max((d_{\mathrm{left}}(v_0)-d_{\mathrm{right}}(v_0))/2,0)$
        \Else
            \State \Return $g_{\mathrm{STCG}}+\max((d_{\mathrm{right}}(v_0)-d_{\mathrm{left}}(v_0))/2,0)$
        \EndIf
    \end{algorithmic}
\end{algorithm}

\begin{algorithm}
    \caption{Path-selected STCG from the right boundary}
    \label{alg_psSTCG}
    \begin{algorithmic}[1]
        \Require A window graph $G=(\mathcal{V},\mathcal{E})$ with edge weight $w$, commit boundary nodes $v_{\mathrm{com},R}$ and $v_{\mathrm{com},L}$, buffer boundary nodes $v_{\mathrm{buf},R}$ and $v_{\mathrm{buf},L}$, and virtual boundary node $v_{\mathrm{vir}}$, distance $d_{\mathrm{left}}$ and $d_{\mathrm{right}}$ from each node to the left and right boundaries in the global decoding graph, and syndrome measurement outcome $\sigma$.
        \Ensure Path-selected STCG from the right boundary
        \State {Compute the minimum weight error $E_{\mathrm{min}}$ and its weight $w_{\mathrm{min}}$ in $G$ with input $\sigma$.}
        \For {Each $v\in N(v_{\mathrm{buf},R})\cup N(v_{\mathrm{buf},L})$}
            \If {$v\in N(v_{\mathrm{buf},R})$}
                \State {$e\gets(v,v_{\mathrm{buf},R})$}
            \Else
                \State {$e\gets(v,v_{\mathrm{buf},L})$}
            \EndIf
            \If {$ E_{\mathrm{min}}(e)=1$}
                \State {Add a new node $u_v$ to $\mathcal{V}$ and a new edge $(v,u_v)$ to $\mathcal{E}$ with weight $w(e)$.}
                \State {$\sigma(u_v)\gets 1$}
            \EndIf
            \State {Eliminate $e$ from $\mathcal{E}$.}
        \EndFor
        \For {Each $v\in N(v_{\mathrm{vir}})$}
            \State {$e\gets(v,v_{\mathrm{vir}})$}
            \State Let $e'=(v,v_0)$ be the edge in the global decoding graph corresponding to $e$, where $v_0$ is the node that was replaced by $v_{\mathrm{vir}}$.
            \State {$w(e)\leftarrow w(e)+\max((d_{\mathrm{left}}(v_0)-d_{\mathrm{right}}(v_0))/2,0)$}
        \EndFor
        \For {Each defect $u$ matched to $v_{\mathrm{vir}}$ in $E_{\mathrm{min}}$}
            \For {Each $v\in N(u)$}
                \State {$e\gets(u,v)$}
                \State {$w(e)\leftarrow w(e)+\max((d_{\mathrm{left}}(u)-d_{\mathrm{right}}(u))/2,0)$}
            \EndFor
            \If {$d_{\mathrm{left}}(u)>d_{\mathrm{right}}(u)$}
                \State {Add edge $e=(u,v_{\mathrm{vir}})$ with weight equal to the distance from $u$ to $v_{\mathrm{vir}}$.}
            \EndIf
        \EndFor
        \State {$\sigma(v_{\mathrm{com},R})\gets 1 \oplus \bigoplus_{v\in N(v_{\mathrm{com},R})}E_{\mathrm{min}}((v,v_{\mathrm{com},R}))$}
        \State Compute the minimum weight error $E_{\mathrm{alt,R}}$ and its weight $w'_{\mathrm{alt,R}}$ in $G$ with input $\sigma$.
        \State \Return $w'_{\mathrm{alt,R}}-w_{\mathrm{min}}$
    \end{algorithmic}
\end{algorithm}

\section{Numerical Analysis}
\label{sec_numerical_analysis}
In this section, we present numerical results for the proposed gaps and the adaptive window decoding based on them.

\subsection{Repetition Code with Phenomenological Noise}
\subsubsection{Properties of the Proposed Gaps}
We first compute the STCG, distance-shifted STCG, and path-selected STCG in sliding window decoding without switching, with the commit and buffer region sizes set to a common value $r_{\mathrm{com}}=r_{\mathrm{buf}}=\lfloor d/2\rfloor$. We define the window-induced logical error rate as the probability of events in which window decoding leads to a logical error while global decoding does not. We simulate the memory experiment on the repetition code under phenomenological noise over $5d$ rounds. The error rate is taken to be uniform across the decoding graph, with $p=0.05$ on each edge and unit weight assigned to every edge. We used PyMatching~\cite{Higgott2022PyMatching}, an efficient implementation of MWPM, as the decoder. Unless otherwise stated, all subsequent data are obtained from $10^7$ Monte Carlo samples.

\begin{figure*}
    \centering
  \begin{tabular}{ll}
{\normalsize (a)} & {\normalsize (b)}  \\
   \includegraphics[width=0.5\linewidth]{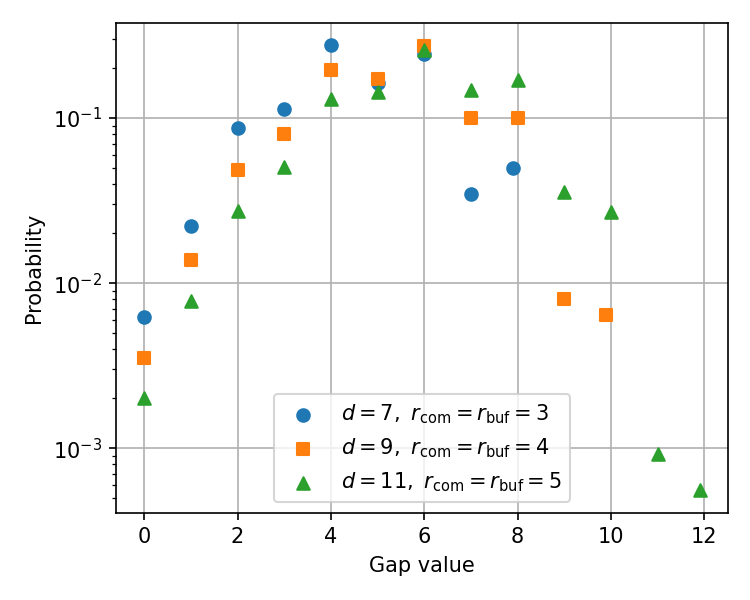}&
   \includegraphics[width=0.5\linewidth]{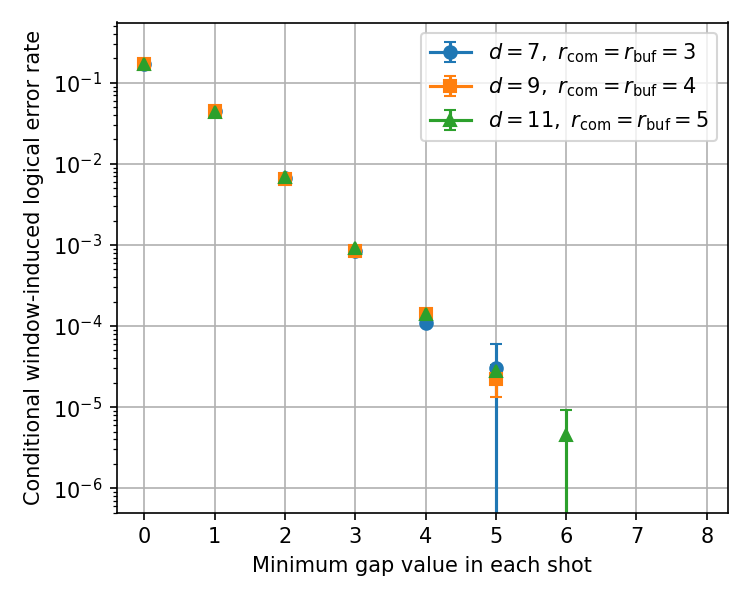}
   \\
  \end{tabular}
    \caption{The path-selected STCG computed for each window in sliding window decoding of the repetition code under the phenomenological noise model. (a) Probability distribution of the gap. (b) Window-induced logical error rate conditioned on the gap value. The physical error rate is $p=0.05$, and syndrome measurements are repeated over $5d$ rounds.}
    \label{fig_psSTCG_property_rep_code}
\end{figure*}

Figure~\ref{fig_psSTCG_property_rep_code} (a) and (b) show the probability distribution of the gap and the window-induced logical error rate conditioned on the minimum gap value over all windows in a shot for the path-selected STCG, respectively. Analogous results for the other two gaps are presented in Appendix~\ref{sec_appendix_other_gaps_properties}. The probability distribution shifts toward larger gap values as $d$ and $r_{\mathrm{buf}}$ increase, indicating that small gap values become less likely. In addition, the window-induced logical error rate conditioned on the gap decreases exponentially with the gap. These behaviors are in approximate agreement with those of the complementary gap~\cite{Gidney2025yoked,toshio2025decoder,Akahoshi2025timelike}.

\subsubsection{Performance of Switching}
We next apply the adaptive sliding window decoding based on each of the three proposed gaps to the repetition code of the code distance $d=13$ under the phenomenological noise model with the physical error rate $p=0.025$. Here $r_{\mathrm{buf}}$ denotes the default buffer size used before switching is invoked.

\begin{figure*}
    \centering
  \begin{tabular}{ll}
{\normalsize (a)} & {\normalsize (b)}  \\
   \includegraphics[width=0.5\linewidth]{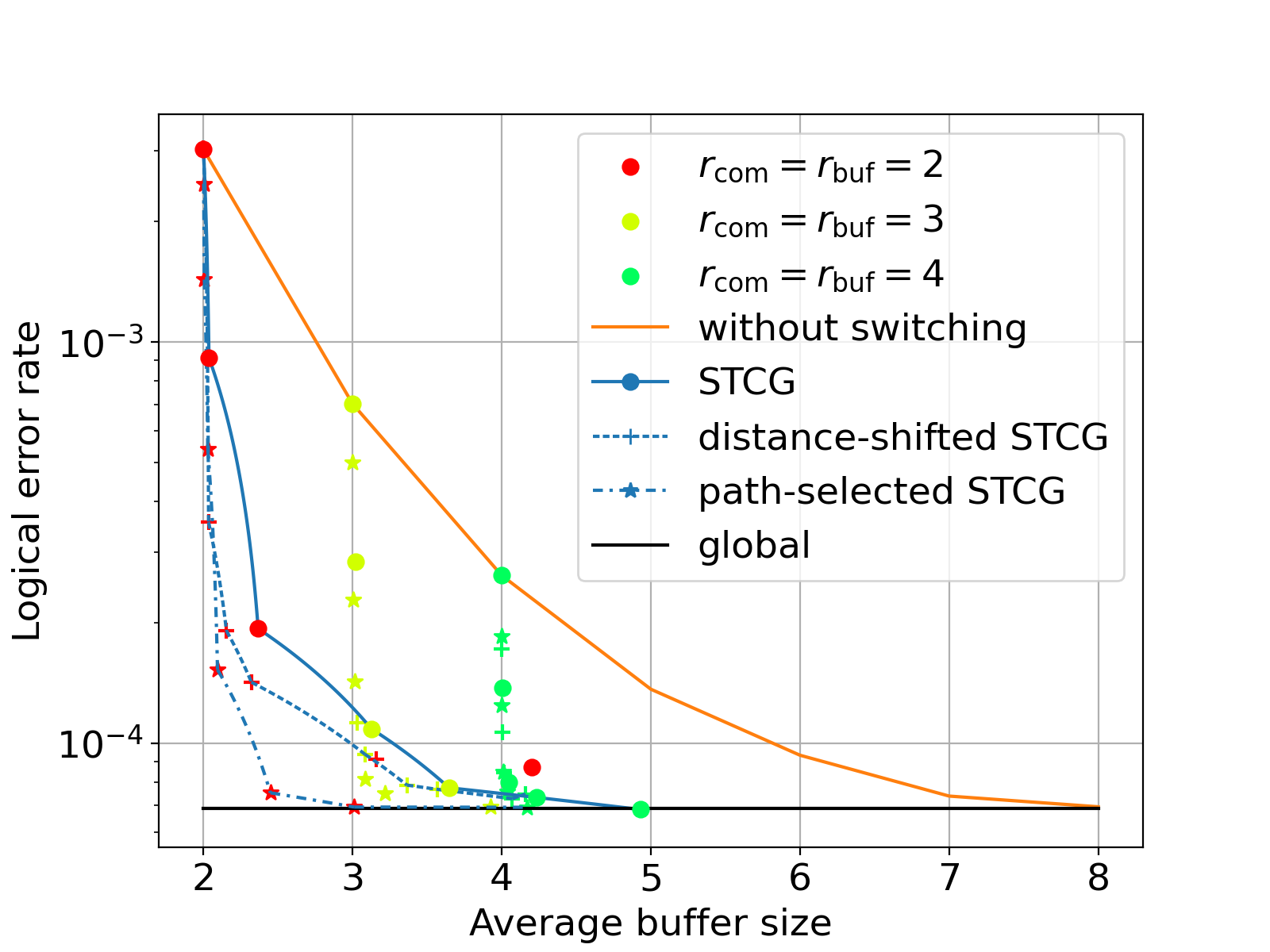}&
   \includegraphics[width=0.5\linewidth]{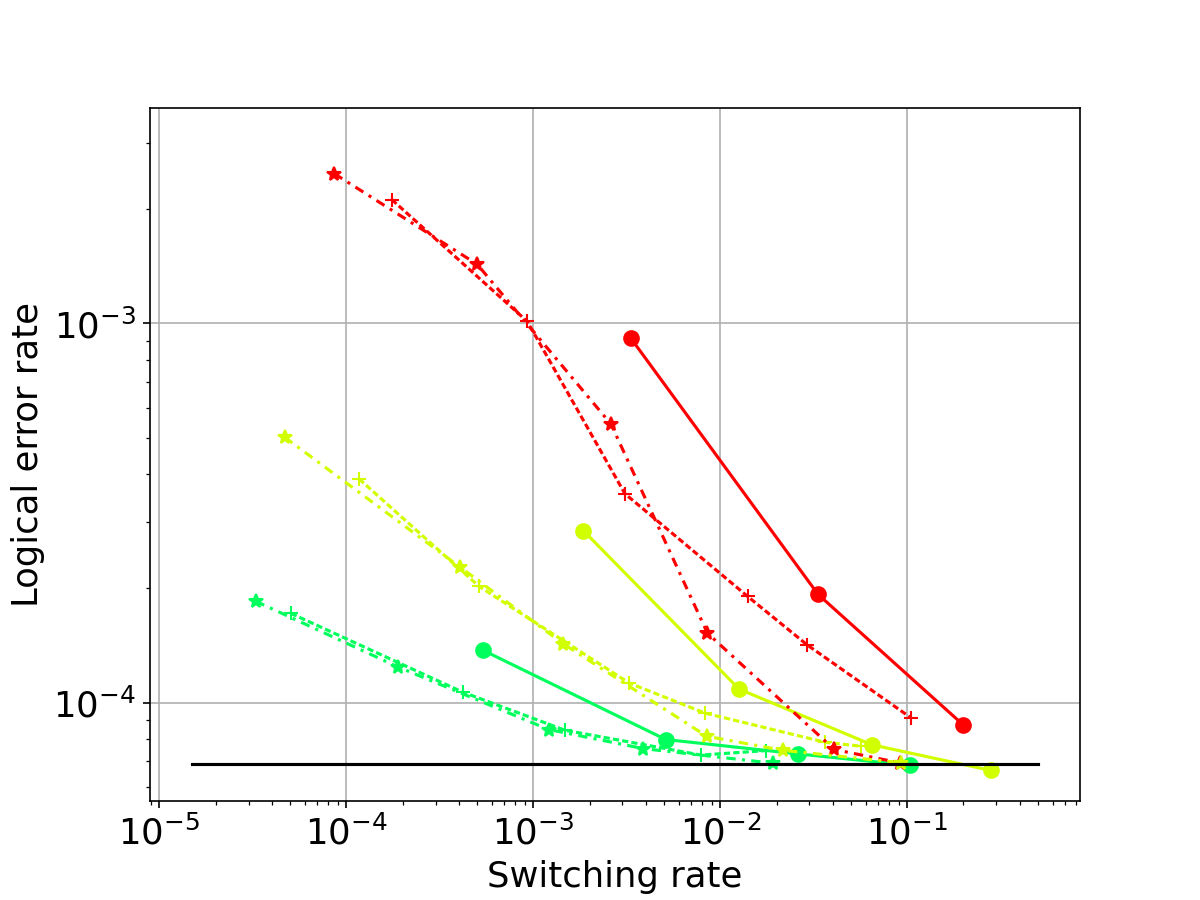}
   \\
  \end{tabular}
    \caption{Adaptive sliding window decoding applied to the repetition code under the phenomenological noise model. (a) Logical error rate as a function of the average buffer size. The points achieving the best trade-offs are connected by lines for each gap. (b) Logical error rate as a function of the switching rate. The code distance is $d=13$, the physical error rate is $p=0.025$, and syndrome measurements are repeated over $5d$ rounds. The gap values are discrete because the edge weights are uniform, which results in the discrete appearance of the plot.}
    \label{fig_result_switching_rep}
\end{figure*}

Figure~\ref{fig_result_switching_rep} (a) shows the logical error rate as a function of the average buffer size. For $d=13$, the average buffer size of approximately $2.5$ is sufficient to attain a logical error rate comparable to the global decoder, while a buffer size of $r_{\mathrm{buf}}=7$ is required to reach the same level of performance without switching.

Figure~\ref{fig_result_switching_rep} (b) shows the trade-off between the logical error rate and the switching rate. For small $r_{\mathrm{buf}}$, the performance increases progressively from the STCG to the distance-shifted STCG, and further to the path-selected STCG. As $r_{\mathrm{buf}}$ increases, however, the distance-shifted STCG and the path-selected STCG become comparable in some regimes. When the path-selected STCG is used, the proposed scheme achieves the same accuracy as the global decoder at a switching rate of $10^{-3}$ with $r_{\mathrm{buf}} = 4$, and at a switching rate of $10^{-2}$ with $r_{\mathrm{buf}} = 3$.

\subsection{Surface Code with Circuit-level Noise}
For circuit-level noise, we focus on the STCG and the path-selected STCG. As mentioned above, the STCG is the simplest of the three gaps, and the path-selected STCG yields the best switching performance for the repetition code; the distance-shifted STCG, lying in between, is omitted for simplicity.

\subsubsection{Properties of Proposed Gaps}
\begin{figure*}
    \centering
  \begin{tabular}{ll}
{\normalsize (a)} & {\normalsize (b)}  \\
   \includegraphics[width=0.5\linewidth]{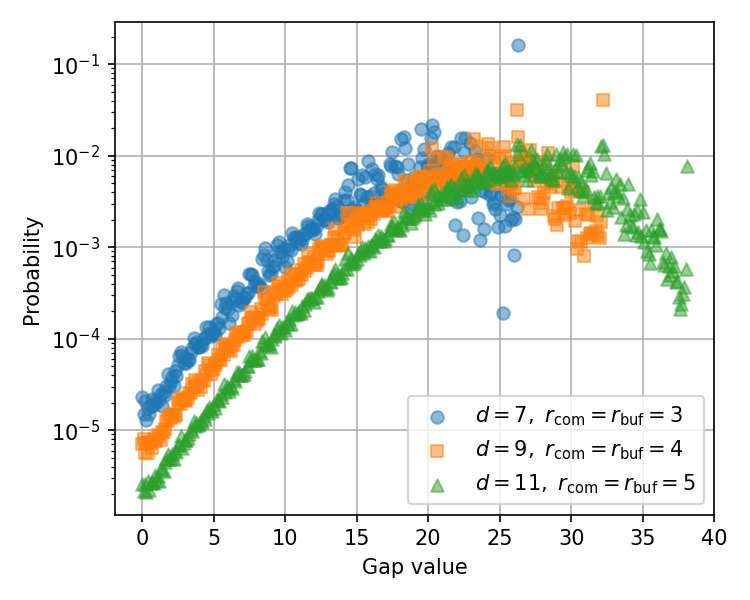}&
   \includegraphics[width=0.5\linewidth]{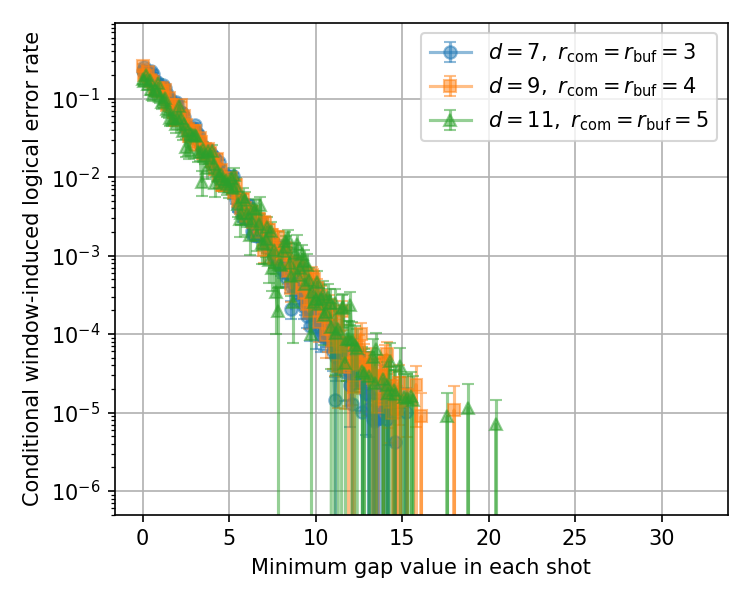}
   \\
  \end{tabular}
    \caption{The path-selected STCG computed for each window in sliding window decoding of the surface code under the circuit-level noise model. (a) Probability distribution of the gap. (b) Window-induced logical error rate conditioned on the gap value. The physical error rate is $p=0.0025$, and syndrome measurements are repeated over $5d$ rounds.}
    \label{fig_psSTCG_property_surface_code}
\end{figure*}

Figure~\ref{fig_psSTCG_property_surface_code} presents the probability distribution of the path-selected STCG and the window-induced logical error rate conditioned on the per-shot minimum gap, evaluated for sliding window decoding on the surface code under uniform circuit-level noise $p=0.0025$. As in the case of the repetition code, the probability distribution shifts toward larger gap values as $d$ and $r_{\mathrm{buf}}$ increase, and the conditional window-induced logical error rate decreases exponentially with the gap. Circuit generation and syndrome sampling were performed using Stim~\cite{Gidney2021stim}.

\subsubsection{Performance of Switching}
\begin{figure*}
    \centering
  \begin{tabular}{ll}
{\normalsize (a)} & {\normalsize (b)}  \\
   \includegraphics[width=0.5\linewidth]{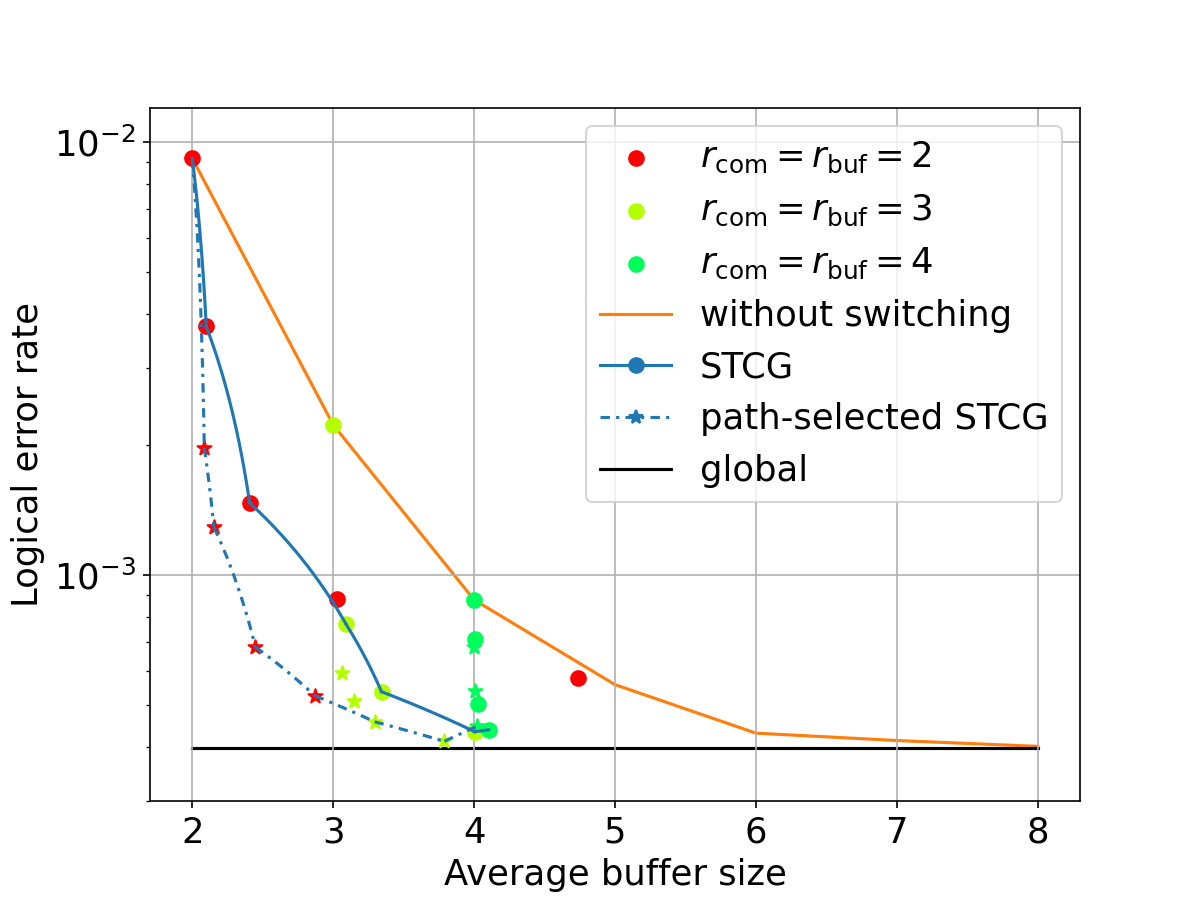}&
   \includegraphics[width=0.5\linewidth]{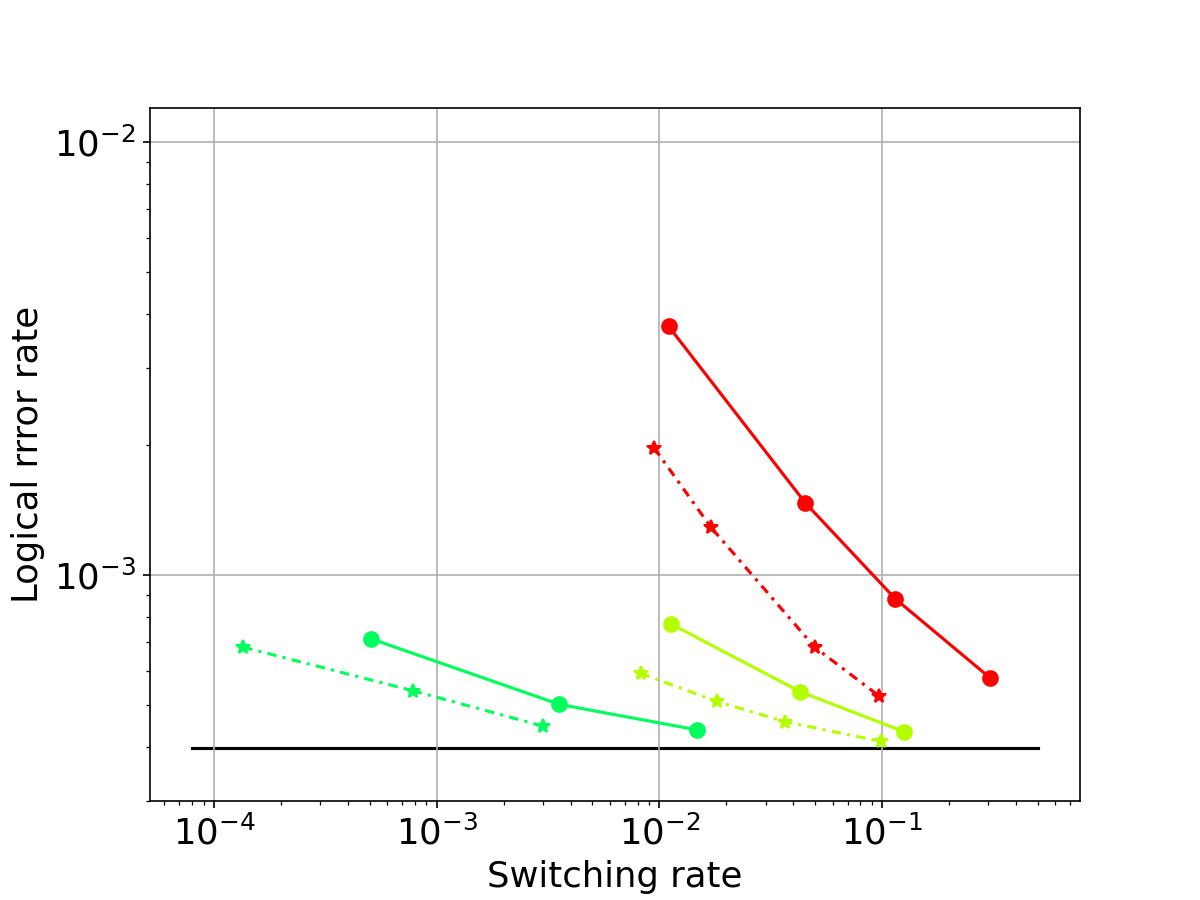}
   \\
  \end{tabular}
    \caption{Adaptive sliding window decoding applied to the surface code under the circuit-level noise model. (a) Logical error rate as a function of the average buffer size. The points achieving the best trade-offs are connected by lines for each gap. (b) Logical error rate as a function of the switching rate. The code distance is $d=11$, the physical error rate is $p=0.0025$, and syndrome measurements are repeated over $5d$ rounds.}
    \label{fig_result_switching_surface}
\end{figure*}

We also perform adaptive sliding window decoding on the surface code of the code distance $d=11$ under circuit-level noise with the physical error rate $p=0.0025$. Figure~\ref{fig_result_switching_surface} (a) shows the logical error rate as a function of the average buffer size, and Fig.~\ref{fig_result_switching_surface} (b) shows the logical error rate as a function of the switching rate. With switching, the average buffer size is reduced from 6 to approximately 3.7 while maintaining a comparable logical error rate. In addition, the proposed scheme achieves the same accuracy as the global decoder at a switching rate of $10^{-2}$--$10^{-3}$ with $r_{\mathrm{buf}} = 4$, and at a switching rate of $10^{-1}$--$10^{-2}$ with $r_{\mathrm{buf}} = 3$. However, the reduction is more limited than in the case of the repetition code under phenomenological noise. This behavior may be attributable to two factors: the increased dimensionality of the decoding graph---from two dimensions for the repetition code to three for the surface code---and the presence of diagonal edges arising from hook errors.
\subsection{Extrapolation to Low Physical Error Rates}

In the regime of low physical error rate with large code distance and buffer size, both the logical error rate and the switching rate become extremely small, which poses a technical challenge for Monte Carlo simulation. Here, we present a procedure for extrapolating data obtained at high physical error rates to the low-error-rate regime. We focus on the path-selected STCG, which achieves the best performance. We first describe the concrete extrapolation procedure under several assumptions, and then verify that these assumptions hold.

\begin{enumerate}
    \item Obtain the gap distribution and the conditional window-induced logical error rate for a single window:
    \begin{enumerate}
        \item Obtain the gap distribution at the target physical error rate by direct sampling.
        \item Obtain the conditional window-induced logical error rate at a high physical error rate,  optionally applying a fit, and use it as a proxy for that at the target physical error rate, based on the assumption that this quantity is universal with respect to the physical error rate.
    \end{enumerate}
    \item Compute the gap distribution and the conditional window-induced logical error rate for the multiple-window setting under the assumption of independence between windows.
    \item Calculate the logical error rate $P_{L}(g_{\mathrm{th}})$ and the switching rate $P_{\mathrm{switch}}(g_{\mathrm{th}})$ as a function of $g_{\mathrm{th}}$ using Eqs.~\eqref{eq_integral_approximation_1} and \eqref{eq_integral_approximation_2}, under the assumption that the approximation introduced therein is valid. Here, $P(g)$ is the gap distribution, $P_{L,\mathrm{global}}$ is the logical error rate of the global decoder, and $P(\mathrm{window\!:\!fail}\,\cap\,\mathrm{global\!:\!success}\mid g)$ is the window-induced logical error rate conditioned on the gap value $g$.
\end{enumerate}

\begin{align}
    &P_{L}(g_{\mathrm{th}}) = P_{L,\mathrm{global}} \notag \\
    & + \int_{g_{\mathrm{th}}}^\infty P(\mathrm{window\!:\!fail}\,\cap\,\mathrm{global\!:\!success}\mid g)\,P(g)\,\d g, \label{eq_integral_approximation_1}\\
    &P_{\mathrm{switch}}(g_{\mathrm{th}}) = \int_{0}^{g_{\mathrm{th}}} P(g)\,\d g. \label{eq_integral_approximation_2}
\end{align}

Note that $P_{L,\mathrm{global}}$ can be estimated by simple extrapolation using the fitting formula $P_L \propto (p/p_{\mathrm{th}})^{(d+1)/2}$, where $p_{\mathrm{th}}$ is the error threshold of the surface code, if necessary~\cite{Fowler2012}.

\begin{figure}
    \centering\includegraphics[width=\linewidth]{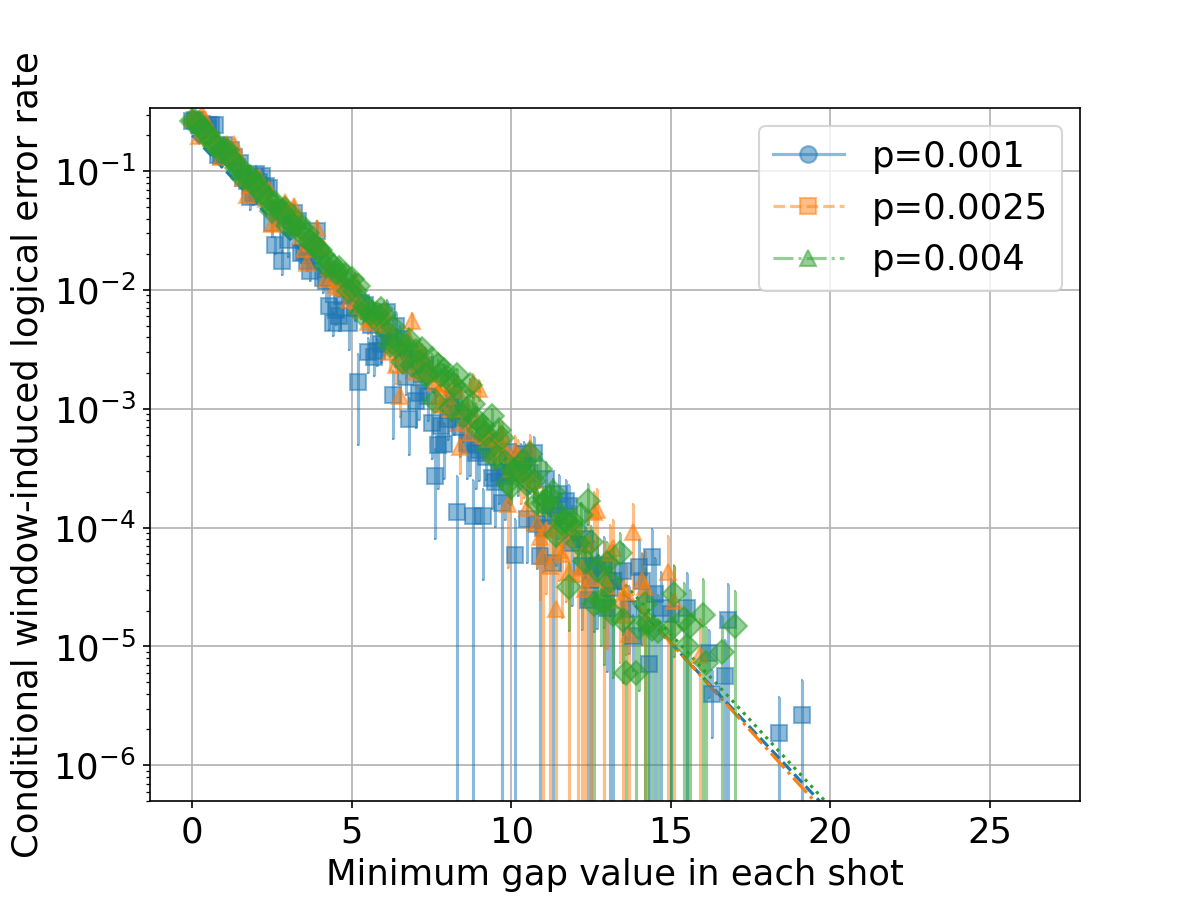}
    \caption{Conditional window-induced logical error rate for various physical error rates. The data points lie on nearly the same line regardless of the physical error rate. $d=7$, $r_{\mathrm{com}}=7$, $r_{\mathrm{buf}}=2$. For $p=0.001$, $10^8$ samples were used.}
    \label{fig_universality_against_physical_error_rate}
\end{figure}

The first assumption is that the conditional window-induced logical error rate is invariant with respect to the physical error rate. Figure~\ref{fig_universality_against_physical_error_rate} shows this quantity for several physical error rates: even as the physical error rate varies, the data points lie on nearly the same fitting curve.

\begin{figure*}
    \centering
  \begin{tabular}{ll}
{\normalsize (a)} & {\normalsize (b)}  \\
   \includegraphics[width=0.5\linewidth]{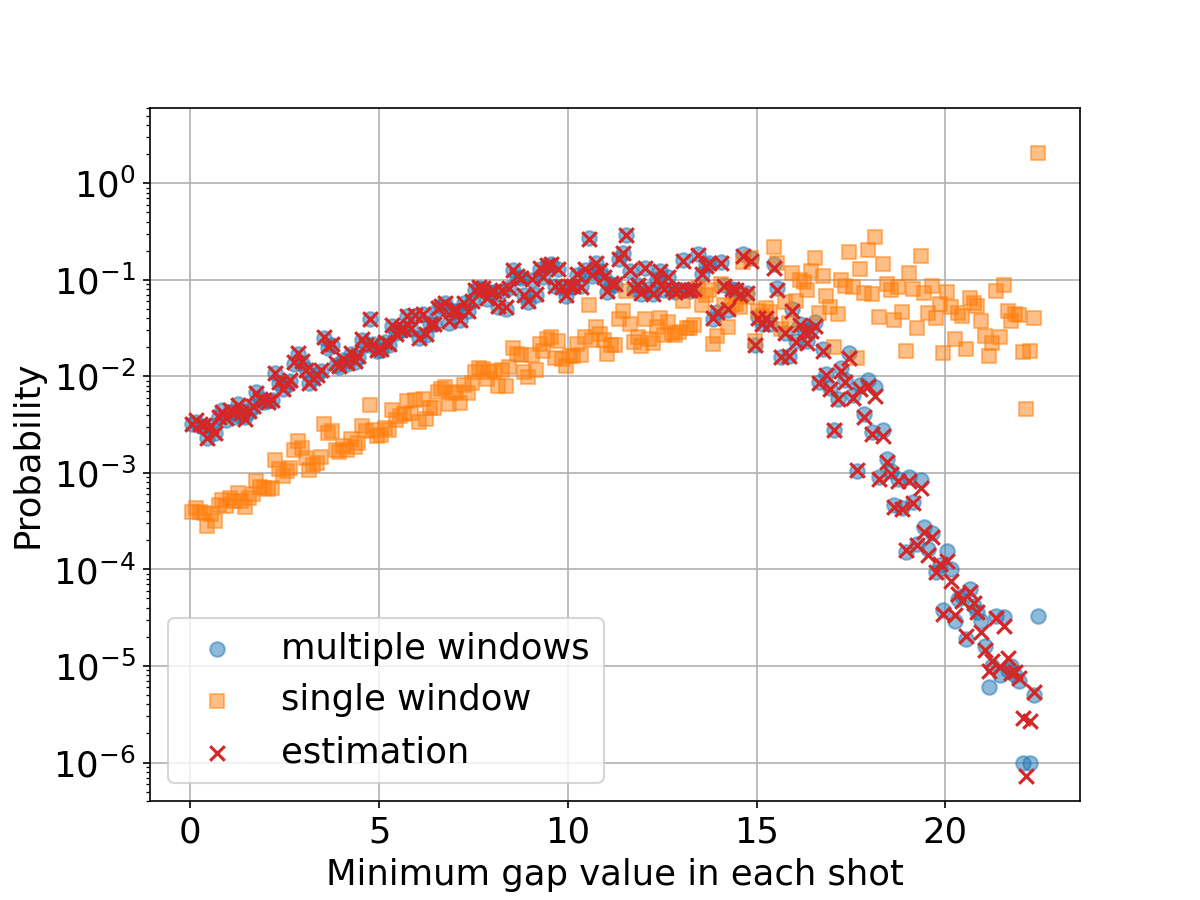}&
   \includegraphics[width=0.5\linewidth]{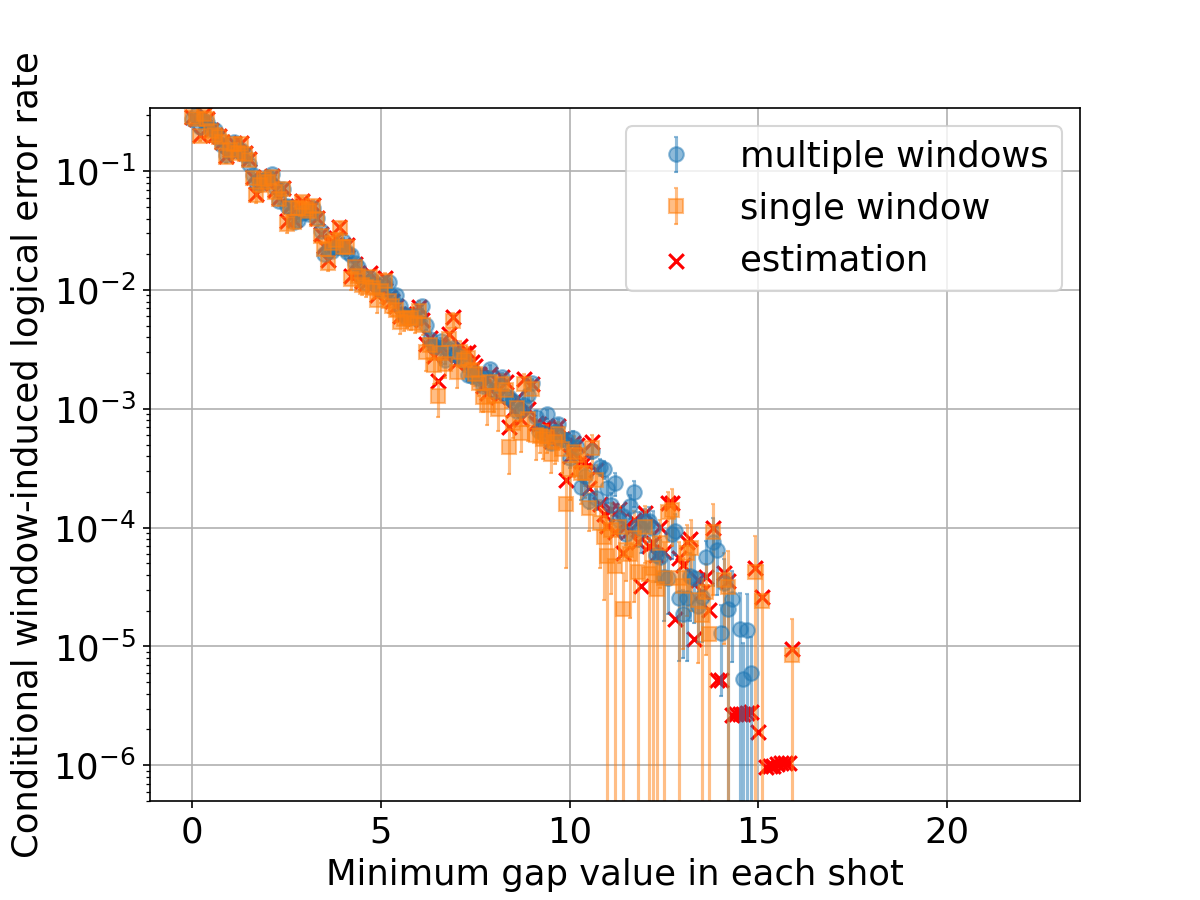}
   \\
  \end{tabular}
    \caption{(a) Gap distribution in a single window and the per-shot minimum-gap distribution in multiple windows. (b) Conditional window-induced logical error rates for the single-window setting and the multiple-window setting. The cross markers represent the prediction of the multiple-window data from the single-window data under the independence assumption. In both (a) and (b), the prediction agrees well with the actual data. In the multiple-window setting, the global decoding graph contains $n=8$ windows. $d=7$, $r_{\mathrm{com}}=7$, $r_{\mathrm{buf}}=2$, $p=0.0025$.}
    \label{fig_psSTCG_property_mutiple_single_comparison}
\end{figure*}

The second assumption is that the gap distribution and the conditional window-induced logical error rate are independent between windows. To examine this, we compare two settings: one in which only a single small-buffer window exists within the global decoding graph, and one in which multiple small-buffer windows are present. In the former, only the single window can cause a window-induced logical error. We further assume $r_{\mathrm{com}} = d$ so that the windows are sufficiently separated from one another. Figure~\ref{fig_psSTCG_property_mutiple_single_comparison} (a) shows the gap distribution in the single-window setting and the per-shot minimum-gap distribution in the multiple-window setting. The cross markers represent the per-shot minimum-gap distribution predicted from the single-window data under the independence assumption. The prediction agrees well with the actual result, supporting the independence of the gap distributions across windows.

Figure~\ref{fig_psSTCG_property_mutiple_single_comparison} (b) shows the conditional window-induced logical error rates $P(\mathrm{fail}\mid g)$ for the single-window setting and $P(\mathrm{fail}\mid \min\{g_1,\ldots,g_n\}=g_{\min})$ for the multiple-window setting, where $g_i$ is a gap value obtained in the $i$-th window. The former is conditioned on the gap obtained in a single window, while the latter is conditioned on the minimum gap among multiple windows. For brevity, we abbreviate ``$\mathrm{window\!:\!fail}\,\cap\,\mathrm{global\!:\!success}$'' as ``fail''. Both settings yield almost identical conditional rates, in agreement with an analytical estimate. Assuming independence between windows, $P(\mathrm{fail}\mid \min\{g_1,\ldots,g_n\}=g_{\min})$ can be approximated from the single-window data as
\begin{align} \label{eq_conditional_ler_analytical_estimation}
    &P(\mathrm{fail}\mid \min\{g_1,\ldots,g_n\}=g_{\min}) \notag \\
    &\quad \simeq P(\mathrm{fail}\mid g_{\min}) + (n-1)\,\frac{\displaystyle\int_{g_{\min}}^\infty P(\mathrm{fail}\,\cap\,g)\,\d g}{\displaystyle\int_{g_{\min}}^\infty P(g)\,\d g},
\end{align}
where $P(g)$ is the gap distribution for a single window, and $P(\mathrm{fail}\,\cap\,g)$ is the joint probability distribution that a window-induced logical error occurs and the gap value $g$ is obtained in a single window. The first term represents the probability of a window-induced logical error in the window with gap $g_{\min}$, and the second term represents the contribution from the remaining $n-1$ windows, each having a gap $g \geq g_{\min}$ and giving rise to a window-induced logical error. The cross markers in Fig.~\ref{fig_psSTCG_property_mutiple_single_comparison} (b) show the estimate obtained from the single-window data via Eq.~\eqref{eq_conditional_ler_analytical_estimation}, which closely matches the single-window result. Both the numerical and analytical results indicate that, even when multiple windows are present, it suffices to consider the window-induced logical error rate conditioned on the minimum gap. Note that the assumption $r_{\mathrm{com}} = d$ is crucial here; if $r_{\mathrm{com}}$ is small, independence between windows is not guaranteed.

The third assumption concerns the approximation used in Eq.~\eqref{eq_integral_approximation_1}. The exact expression for the logical error rate given the gap threshold $g_{\mathrm{th}}$ is
\begin{align}
    P_{L}(g_{\mathrm{th}}) &= P_{L,\mathrm{global}} \notag \\
    & + \int_{g_{\mathrm{th}}}^\infty \big( P(\mathrm{window\!:\!fail}\,\cap\,\mathrm{global\!:\!success}\mid g) \notag \\
    & - P(\mathrm{window\!:\!success}\,\cap\,\mathrm{global\!:\!fail}\mid g) \big)\,P(g)\,\d g, \label{eq_integral_exact}
\end{align}
where $P(\mathrm{window\!:\!success}\,\cap\,\mathrm{global\!:\!fail}\mid g)$ is the conditional probability that the global decoder produces a logical error while the window decoder does not. In practice, this quantity is expected to be much smaller than $P(\mathrm{window\!:\!fail}\,\cap\,\mathrm{global\!:\!success}\mid g)$ and can therefore be neglected.

\begin{figure}
    \centering\includegraphics[width=\linewidth]{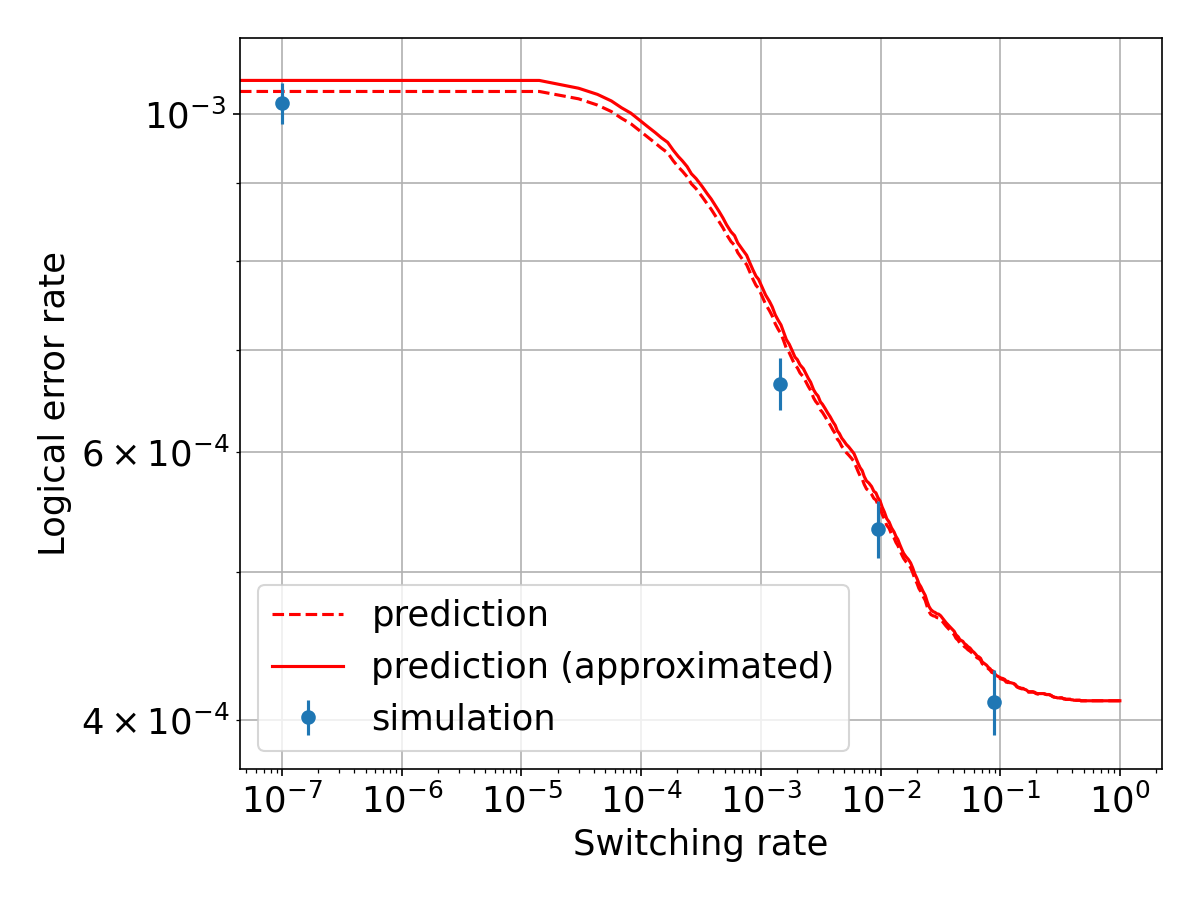}
    \caption{Comparison of the data predicted by Eqs.~\eqref{eq_integral_exact} and \eqref{eq_integral_approximation_2} (dashed line), the data approximately predicted by Eqs.~\eqref{eq_integral_approximation_1} and \eqref{eq_integral_approximation_2} (solid line), and the data obtained from direct simulation. The approximate prediction agrees well with the simulation. $d=11$, $r_{\mathrm{com}}=11$, $r_{\mathrm{buf}}=2$, $p=0.0025$.}
    \label{fig_approximation_validity}
\end{figure}

Figure~\ref{fig_approximation_validity} compares the estimated trade-off between the logical error rate and the switching rate with the actual values. The dashed line shows the exact estimate based on Eqs.~\eqref{eq_integral_exact} and \eqref{eq_integral_approximation_2}, while the solid line shows the approximate estimate based on Eqs.~\eqref{eq_integral_approximation_1} and \eqref{eq_integral_approximation_2}. The two estimates nearly coincide and both agree with the actual values, confirming the validity of the approximation.

\begin{figure*}
    \centering
  \begin{tabular}{ll}
{\normalsize (a)} & {\normalsize (b)}  \\
   \includegraphics[width=0.5\linewidth]{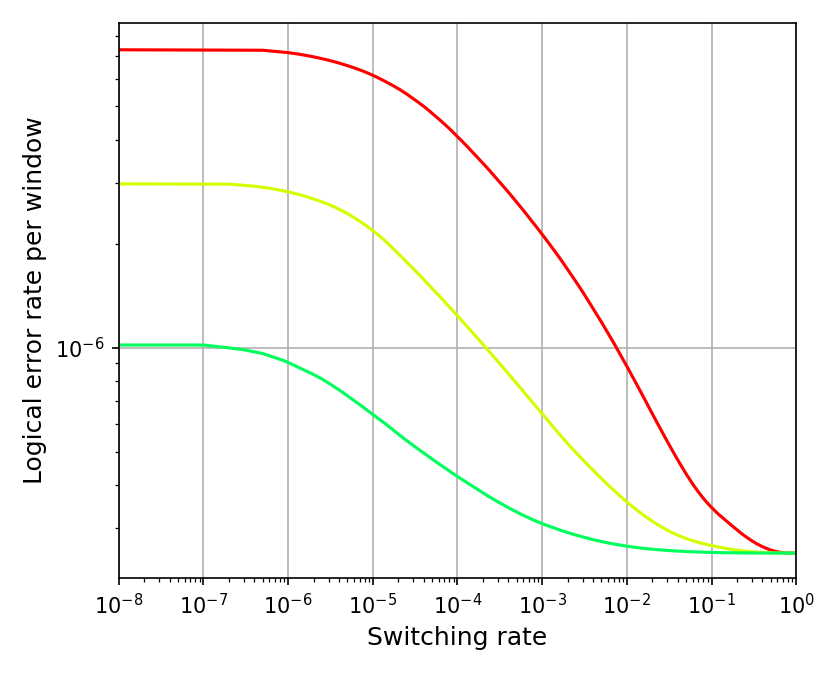}&
   \includegraphics[width=0.5\linewidth]{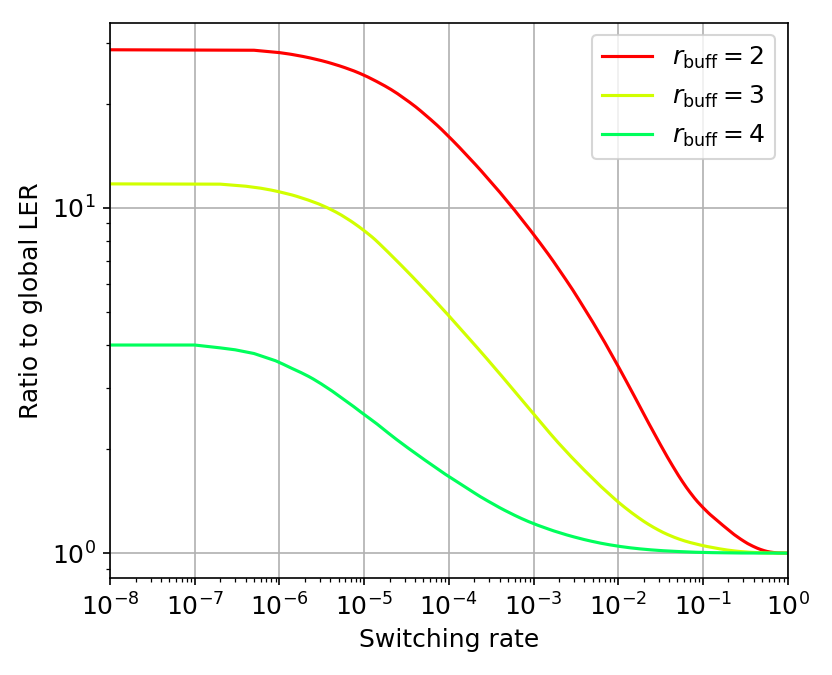}
   \\
  \end{tabular}
    \caption{(a) Logical error rate per window and (b) ratio of the logical error rate to that of the global decoder, at $p=0.001$, extrapolated from the data at $p=0.004$. With $r_{\mathrm{buf}}=4$, the proposed scheme achieves a logical error rate comparable to that of the global decoder at a switching rate of $10^{-3}$--$10^{-4}$; with $r_{\mathrm{buf}}=3$, the same is achieved at a switching rate of $10^{-1}$--$10^{-2}$. $d=11$, $r_{\mathrm{com}}=11$.}
    \label{fig_LER_SR_estimation}
\end{figure*}

We then apply the proposed extrapolation procedure to estimate the logical error rate and the switching rate in the low physical-error-rate regime. We set the code distance to $d=11$ and use the conditional window-induced logical error rate measured at $p = 0.004$ to predict the behavior at $p = 0.001$. The results are shown in Fig.~\ref{fig_LER_SR_estimation}. With $r_{\mathrm{buf}} = 4$, the proposed scheme achieves comparable accuracy as the global decoder at a switching rate of $10^{-3}$--$10^{-4}$.

\section{{Potential Applications}}
\label{sec_discussion}
The method proposed in this work admits several potential applications. The first is the reduction of decoding time. By reducing the average buffer size, we can shorten the average decoding time. One might be concerned that the computation of soft information such as the STCG itself incurs a non-negligible overhead. A possible way to address this overhead is to use recently proposed efficient methods for computing soft information~\cite{Meister2024,Kishi2025}. When Union-Find is used as the inner decoder in combination with these methods, the cost of computing soft information is expected to become nearly negligible. However, how to compute the soft information proposed in this work using such fast methods remains an open question. As for the overhead of redoing the decoding, it can be mitigated by running a decoder with buffer size $d$ in parallel at all times and using its output only when switching is triggered.

The second application is the improvement of the decoding reaction time during non-Clifford gate operations. This goes beyond simply shortening the total decoding time: it also enables decoding to begin earlier. To decode a commit region, the buffer is conventionally required to extend a distance of $d$ beyond it. In the context of gate teleportation, for example, decoding cannot start until $d$ rounds of syndrome measurements have been collected after the teleportation measurement. Our method probabilistically reduces the number of additional syndrome rounds that must be awaited, thereby shortening the average time before decoding can begin.

The third application is switching across distinct hardware platforms. FPGA implementations of decoders are highly desirable for their speed and low latency, but they are constrained by the problem size and become increasingly difficult to realize as the code distance grows~\cite{Liyanage2024FPGA}. This issue becomes especially pronounced in lattice surgery, where buffer regions must be allocated in multiple directions, potentially exceeding the capacity of a single FPGA. A natural strategy in such settings is to offload the large-buffer decoding tasks to a CPU or GPU, which can accommodate larger problem sizes at the cost of higher latency. When the switching rate is sufficiently low, decoding with a small buffer suffices in the majority of cases and the switch to a larger-buffer decoder is invoked only rarely, making this strategy practical. This realizes an architecture analogous to the weak/strong decoder framework proposed in Ref.~\cite{toshio2025decoder}: the weak decoder---here, a window decoder with a small buffer---runs on fast and resource-efficient hardware such as FPGAs or ASICs, while the strong decoder, a window decoder with a buffer of size $d$ (optionally employing a more accurate inner decoder), runs on a CPU or GPU.

\section{Conclusion}
\label{sec_conclusion}
In this paper, we have proposed a method for reducing the window size in sliding-window decoding of the surface code while preventing degradation of the logical error rate. In the proposed scheme, decoding is first performed with a small buffer size, and decoding confidence (soft information) is computed. Only when the confidence is low, the buffer size is reset to the code distance $d$ and decoding is performed again. Since the conventional definitions of decoding confidence are not well-suited to the window decoding, we further proposed a new form of soft information tailored to this setting.

Extensive numerical simulations confirm that the proposed method and the proposed gap exhibit the following properties. First, the conditional logical error rate associated with the proposed gap decreases exponentially with the gap, which is a desirable feature for reducing the switching rate. Second, the proposed method achieves the same logical error rate as the fixed-buffer scheme while using a smaller buffer size, with only a low switching rate. In addition, we presented an extrapolation procedure to the low physical-error-rate regime and showed that, under several reasonable assumptions, the performance at low physical error rates can be predicted from numerical data obtained at high physical error rates. This makes it possible to estimate the performance of the proposed method in the fault-tolerant operating regime, which is otherwise difficult to access by direct Monte Carlo simulation.

The proposed method is particularly well-suited to fault-tolerant quantum computation in which scalable real-time decoding is required. It not only reduces the average decoding time but is also applicable to improving the reaction time during non-Clifford gate execution, and is therefore expected to contribute to the overall throughput of quantum computation.

Several directions remain open for future work. First, while this work assumes MWPM as both the inner and global decoder, it would be valuable to analyze the behavior of the proposed scheme when other decoders---such as Union-Find, belief-matching, or neural network decoders---are used. Second, since various forms of soft information have been proposed for codes beyond the surface code, extending the proposed scheme to other qLDPC codes is a promising research direction~\cite{Lee2025soft,xie2026simple}. Third, evaluating the overhead of gap computation on dedicated hardware (FPGA or ASIC) is an important step toward practical implementation. Through these extensions, the proposed method is expected to contribute to the realization of practical fault-tolerant quantum computation.

\section*{Acknowledgement}
We are grateful to Yutaro Akahoshi, Shinichiro Yamano, Yutaka Hirano, and Yugo Takada for fruitful discussions. K. F. is supported by the MEXT Quantum Leap Flagship Program (MEXT Q-LEAP)
Grant No. JPMXS0120319794, JST COI-NEXT Grant No. JPMJPF2014, and JST
Moonshot R\&D Grant No. JPMJMS2061.

\section*{Data availability}
The data and code that supports the findings of this article is openly available~\cite{source_code}.

\appendix
\section{Properties of the STCG and the Distance-shifted STCG}
\label{sec_appendix_other_gaps_properties}

\begin{figure*}
    \centering
  \begin{tabular}{ll}
{\normalsize (a)} & {\normalsize (b)}  \\
   \includegraphics[width=0.5\linewidth]{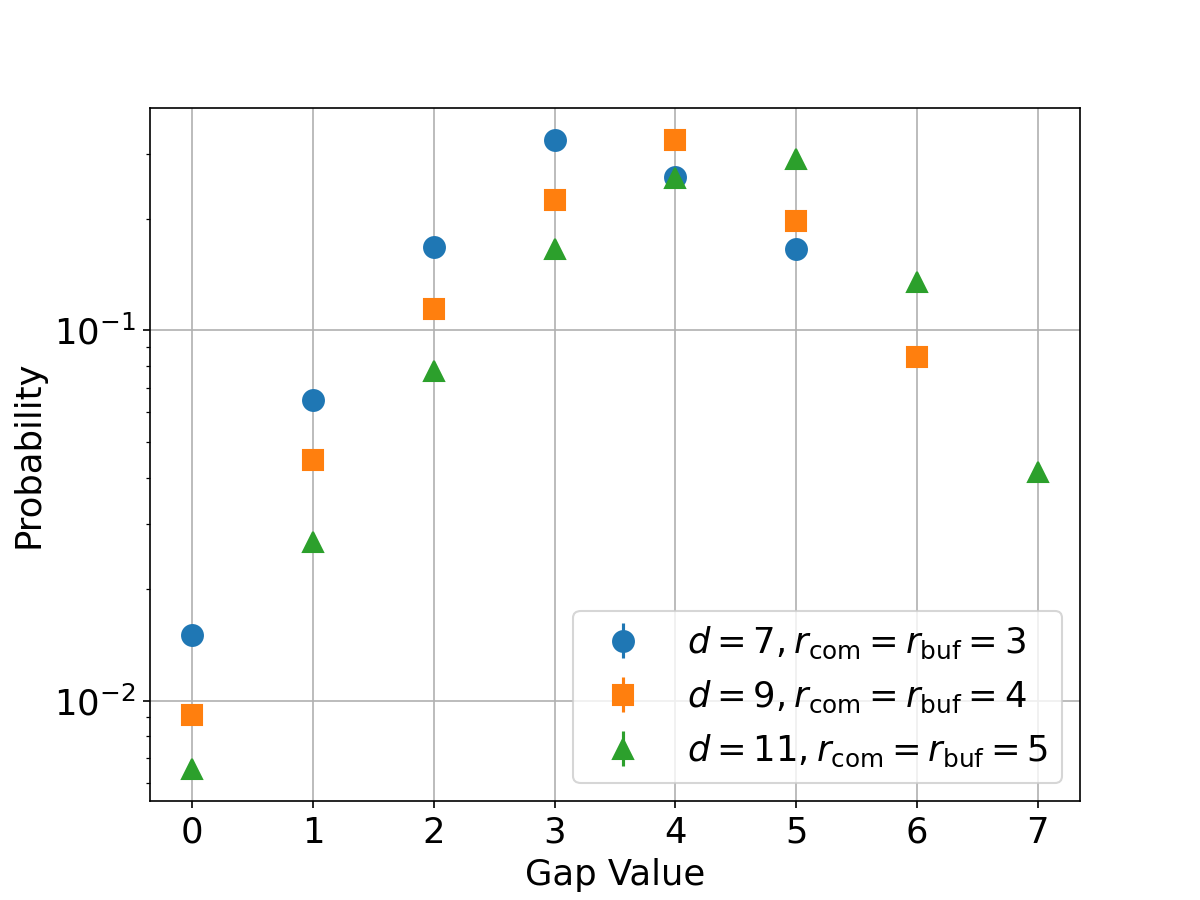}&
   \includegraphics[width=0.5\linewidth]{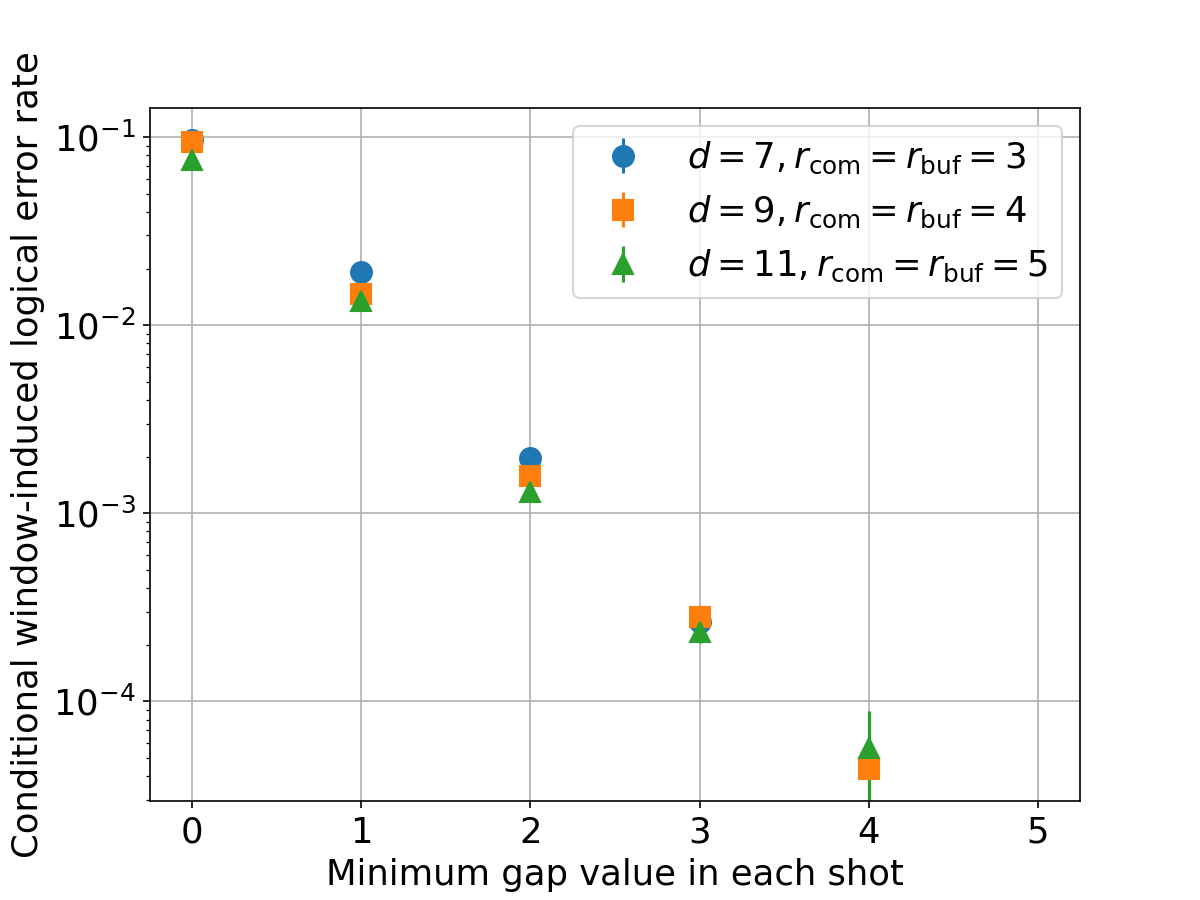}
   \\
  \end{tabular}
    \caption{The STCG computed for each window in sliding window decoding of the repetition code under the phenomenological noise model. (a) Probability distribution of the gap. (b) Window-induced logical error rate conditioned on the gap value.}
    \label{fig_STCG_property_rep_code}
\end{figure*}

\begin{figure*}
    \centering
  \begin{tabular}{ll}
{\normalsize (a)} & {\normalsize (b)}  \\
   \includegraphics[width=0.5\linewidth]{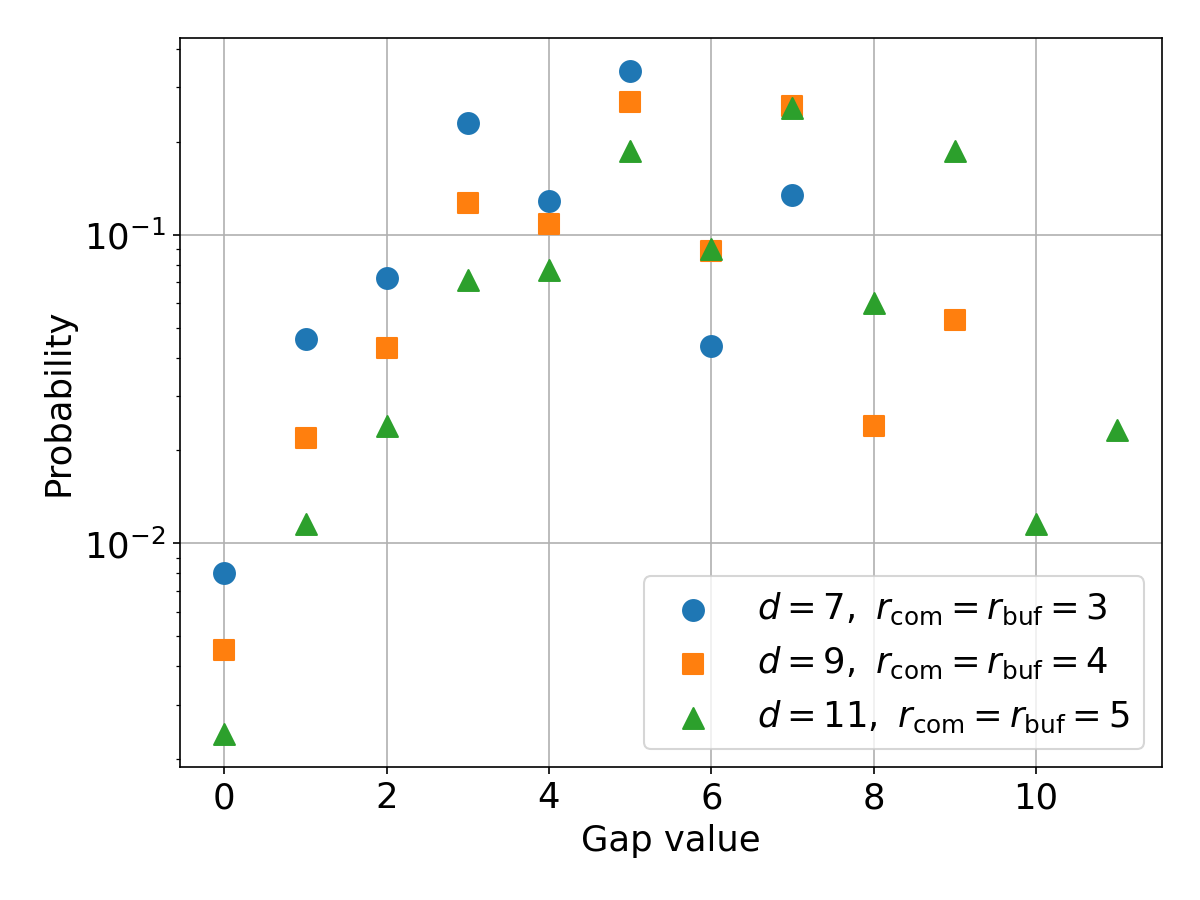}&
   \includegraphics[width=0.5\linewidth]{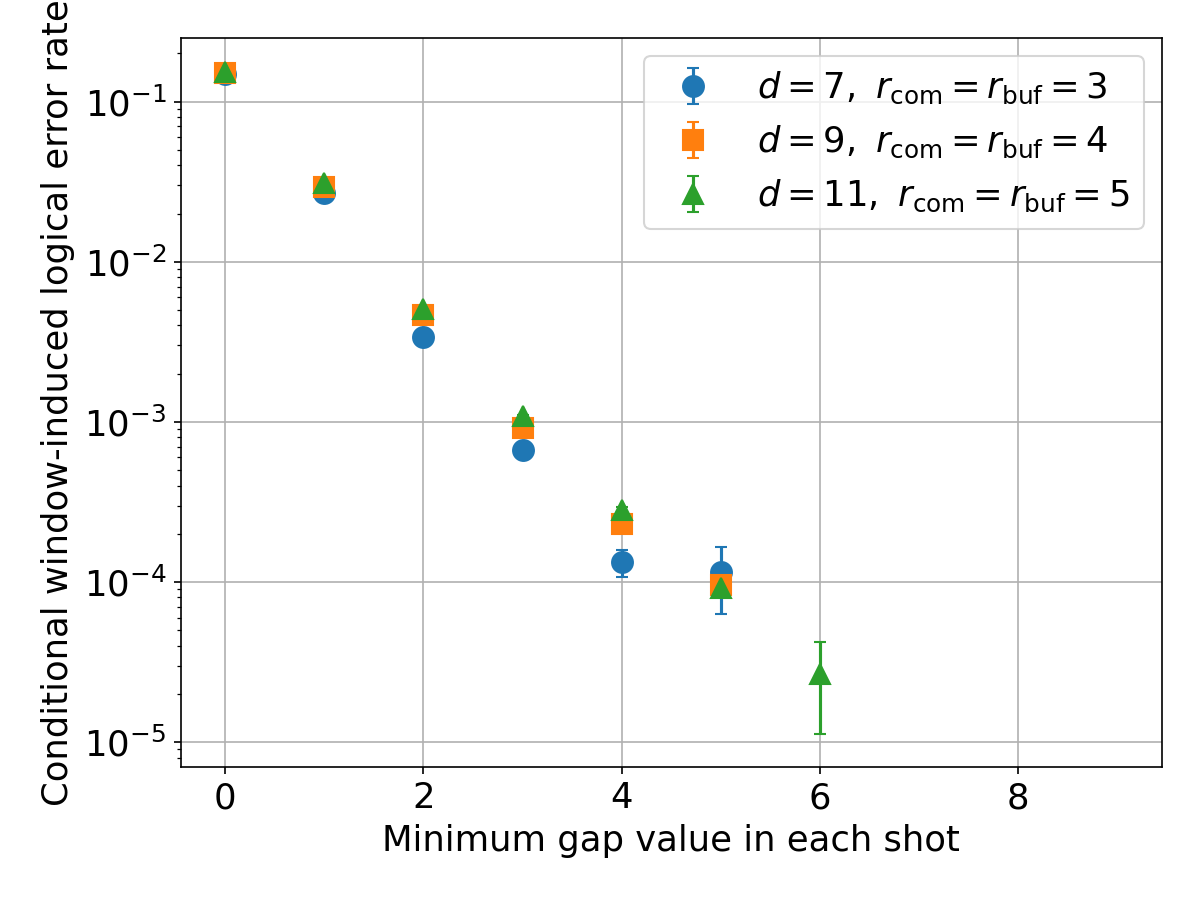}
   \\
  \end{tabular}
    \caption{The distance-shifted STCG computed for each window in sliding window decoding of the repetition code under the phenomenological noise model. (a) Probability distribution of the gap. (b) Window-induced logical error rate conditioned on the gap value.}
    \label{fig_dsSTCG_property_rep_code}
\end{figure*}

\begin{figure*}
    \centering
  \begin{tabular}{ll}
{\normalsize (a)} & {\normalsize (b)}  \\
   \includegraphics[width=0.5\linewidth]{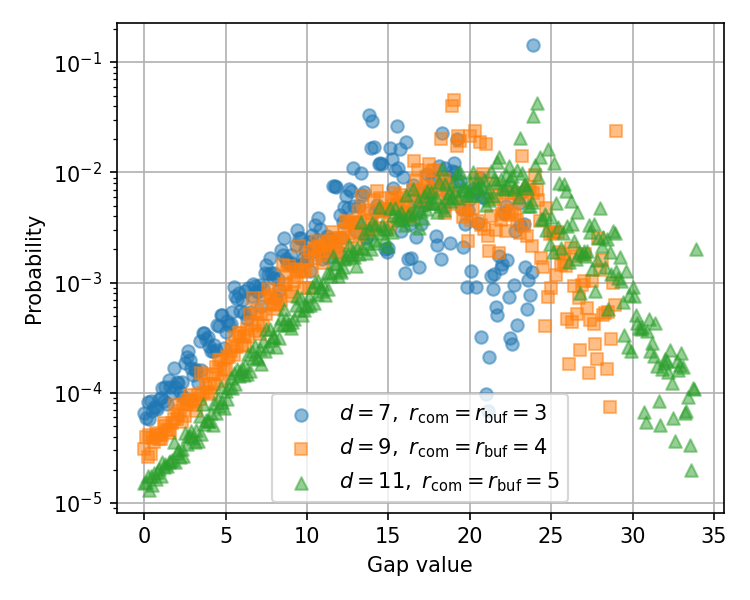}&
   \includegraphics[width=0.5\linewidth]{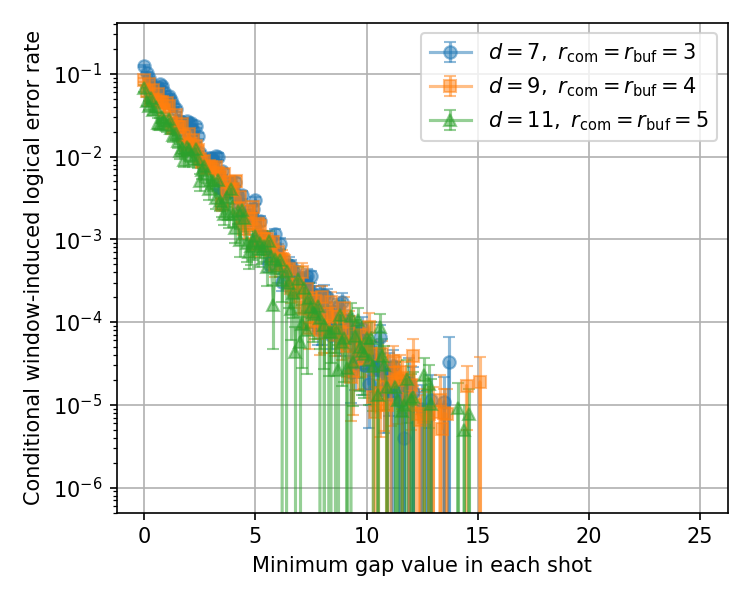}
   \\
  \end{tabular}
    \caption{The STCG computed for each window in sliding window decoding of the surface code under the circuit-level noise model. (a) Probability distribution of the gap. (b) Window-induced logical error rate conditioned on the gap value.}
    \label{fig_STCG_property_surface_code}
\end{figure*}

Figures~\ref{fig_STCG_property_rep_code} and~\ref{fig_dsSTCG_property_rep_code} show the gap distribution and the conditional window-induced logical error rate for the STCG and the distance-shifted STCG, respectively, evaluated on the repetition code. Figure~\ref{fig_STCG_property_surface_code} shows the corresponding results for the STCG on the surface code. The overall shapes of the distributions and the conditional logical error rates are largely similar across the different gaps. The shift of the distribution toward larger gap values with increasing $d$ and $r_{\mathrm{buf}}$ is also common to all of them. Compared with the STCG and the distance-shifted STCG, the conditional window-induced logical error rate of the path-selected STCG exhibits a weaker dependence on $d$. This may reflect the fact that the path-selected STCG more closely captures the physical mechanism underlying logical errors. On the other hand, it is remarkable that the STCG exhibits behavior similar to the path-selected STCG, despite its rather drastic approximation of neglecting $t_{\mathrm{min}}$, as explained in Sec.~\ref{subsubsec_STCG}.

\section{Adaptive Parallel Window Decoding}
\begin{figure}
    \centering
    \includegraphics[width=\linewidth]{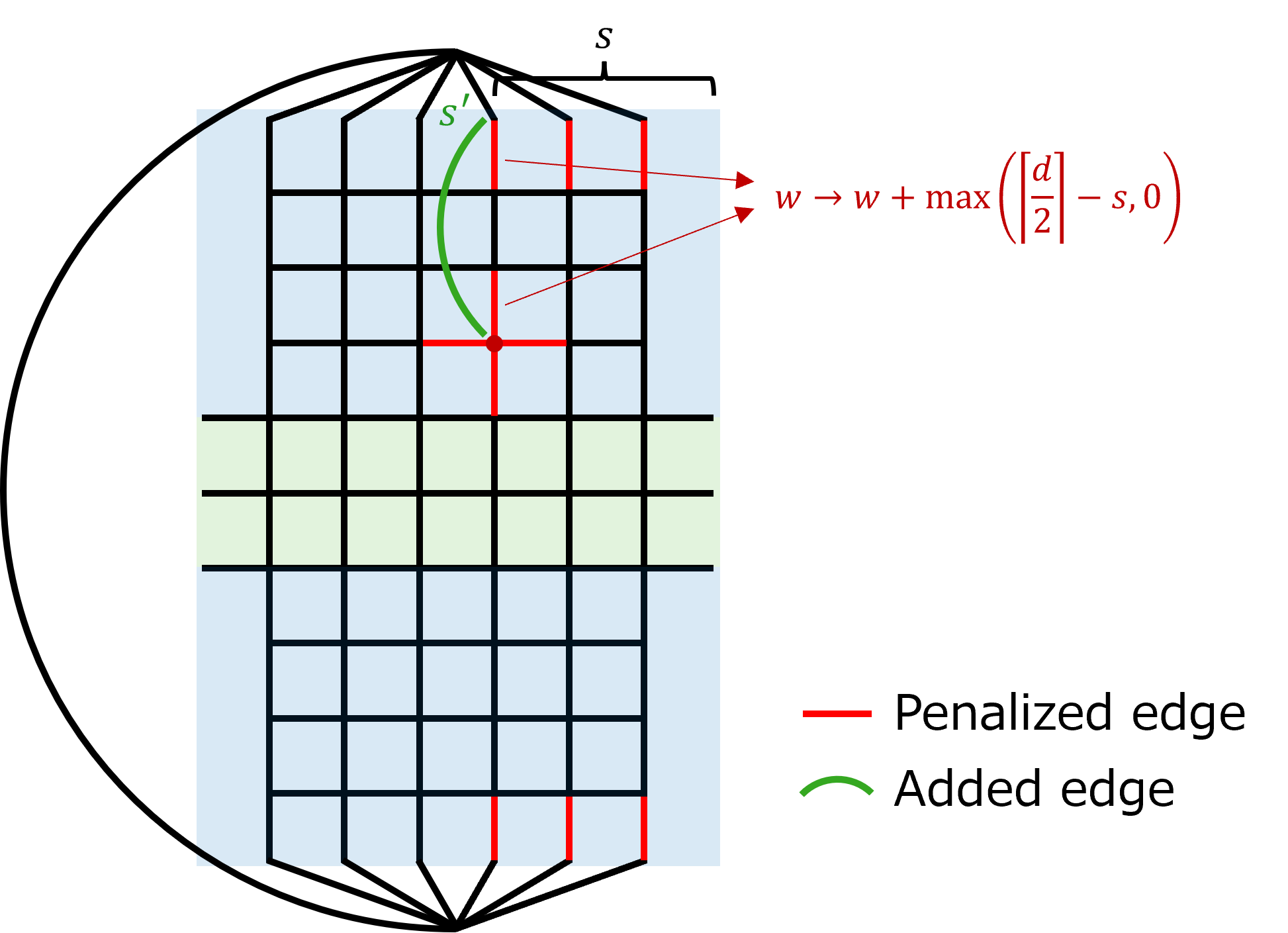}
    \caption{Window used in computing the path-selected STCG under parallel window decoding. For simplicity, the upper and lower virtual boundaries are treated as a single boundary.}
    \label{fig_psSTCG_computation_parallel_window}
\end{figure}

The scheme proposed in this work extends naturally to parallel window decoding in the context of memory experiments. Here, we apply the path-selected STCG to the windows that appear in parallel window decoding. For simplicity, we treat the upper and lower virtual boundary nodes as a single boundary node and output only the smaller of the two gaps associated with the upper and lower virtual boundaries (see Fig.~\ref{fig_psSTCG_computation_parallel_window}). If the two gaps were instead computed separately, switching for the upper and lower virtual boundaries could be invoked independently. Under our simplified implementation, however, when a small gap value $g$ is obtained on one side and indicates that switching is necessary, the other gap is known only to be at least $g$, leaving its switching decision undetermined. Consequently, once switching is triggered, both the upper and lower buffer regions must be extended to size $d$, leading to an excessive increase in the problem size. On the other hand, this implementation reduces the number of gap computations: a naive implementation would require four gap computations in total (possibly in parallel), corresponding to the gaps from the left and right commit boundaries to the upper and lower virtual boundaries, whereas our implementation combines the computations for the upper and lower boundaries, reducing the total to two (possibly in parallel).

\begin{figure*}
    \centering
  \begin{tabular}{ll}
{\normalsize (a)} & {\normalsize (b)}  \\
   \includegraphics[width=0.5\linewidth]{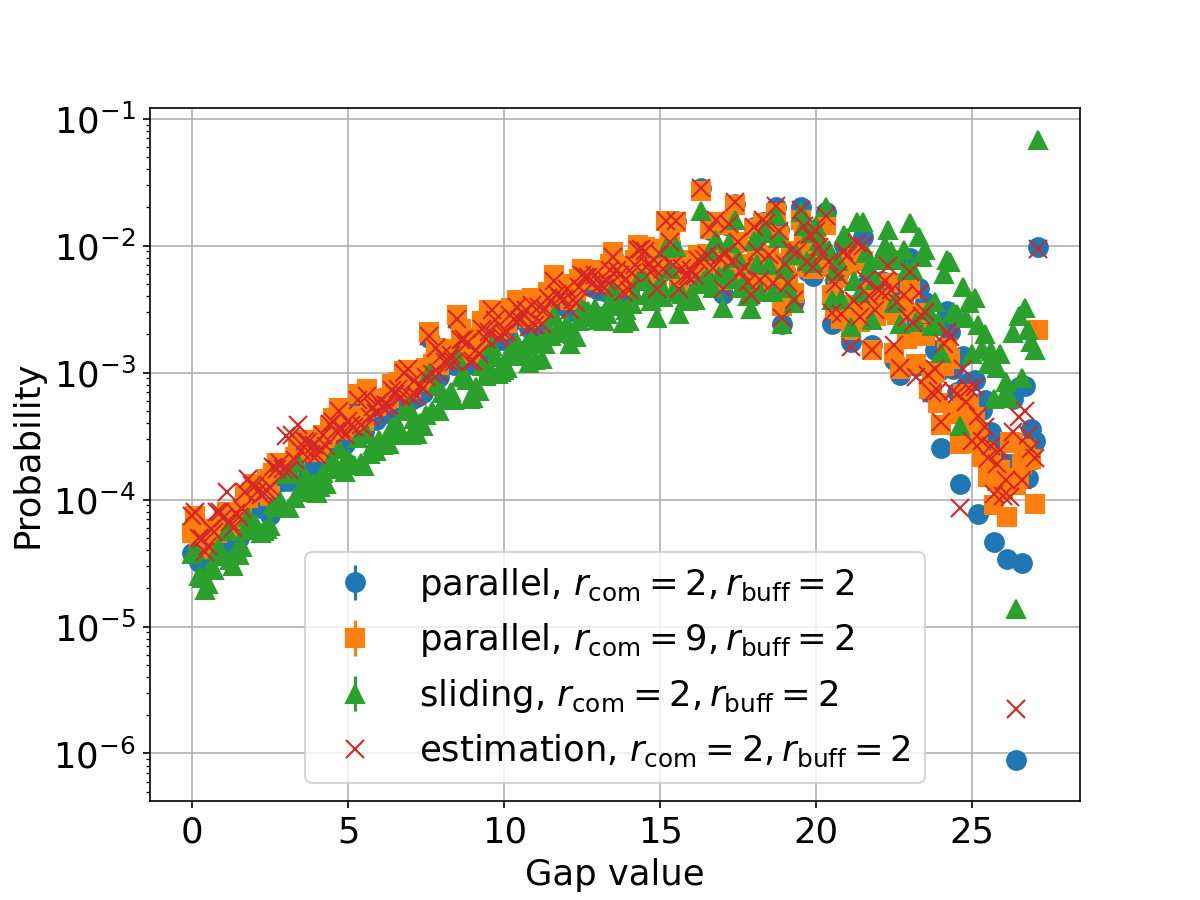}&
   \includegraphics[width=0.5\linewidth]{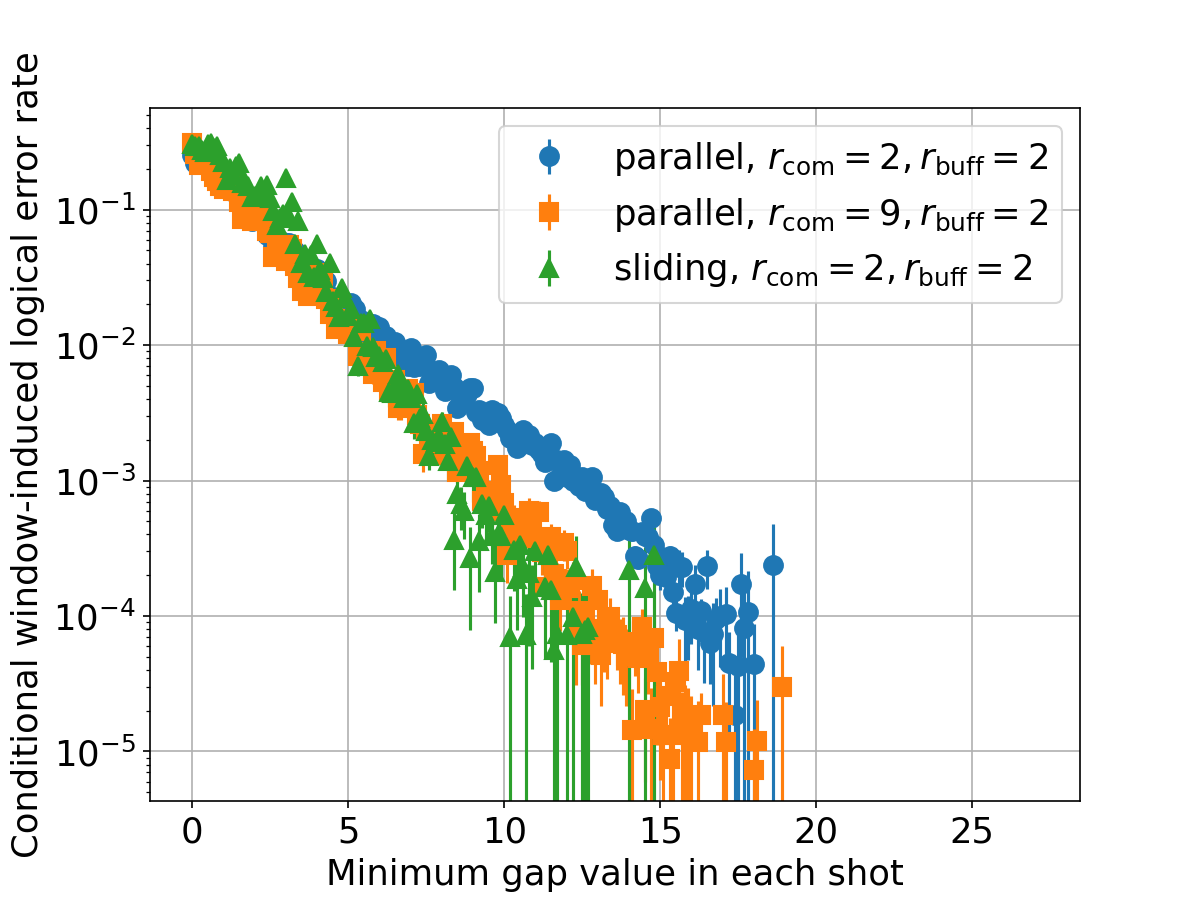}
   \\
  \end{tabular}
    \caption{The path-selected STCG computed for each window in parallel window decoding of the surface code. (a) Probability distribution of the gap. (b) Window-induced logical error rate conditioned on the gap value. $d=9$, $p=0.0025$.}
    \label{fig_psSTCG_property_surface_code_parallel_a}
\end{figure*}

Figure~\ref{fig_psSTCG_property_surface_code_parallel_a} shows the gap distribution and the conditional window-induced logical error rate for the path-selected STCG computed under parallel window decoding on the surface code. We take the code distance to be $d=9$ and consider two cases, $r_{\mathrm{com}}=2$ and $r_{\mathrm{com}}=9$. For comparison, we also include the data for sliding window decoding. As seen in Fig.~\ref{fig_psSTCG_property_surface_code_parallel_a} (a), the gap distributions for $r_{\mathrm{com}}=2$ and $r_{\mathrm{com}}=9$ are nearly indistinguishable. Since the gap computation used here outputs the smaller of the two gaps associated with the upper and lower virtual boundaries, we also compute the distribution of the minimum of two independent samples drawn from the sliding-window data, derived analytically from the empirical distribution; this distribution, shown by cross markers, agrees well with the parallel-window data. This indicates that, as far as the minimum of two gaps is concerned, the sliding-window data are sufficient to predict the behavior of parallel window decoding with adequate accuracy. In contrast, as seen in Fig.~\ref{fig_psSTCG_property_surface_code_parallel_a} (b), the conditional window-induced logical error rate for $r_{\mathrm{com}}=9$ is in reasonable agreement with that of sliding window decoding, whereas for $r_{\mathrm{com}}=2$, it is substantially larger. We attribute this discrepancy to correlations induced by the short commit and buffer regions. Quantitatively characterizing this effect is difficult, and the prediction of the conditional window-induced logical error rate for $r_{\mathrm{com}} < d$, both in parallel and sliding window decoding, remains open. Furthermore, in this work we have focused on memory experiments. Extending the definition of the distance-shifted STCG and the path-selected STCG to windows of general shape, such as those that can arise in lattice surgery, remains an open problem.

\section{Computation of Conventional Complementary Gap} \label{sec_appendix_conventional_cgap}
The complementary gap has found a variety of applications, including post-selection~\cite{Bombin2024,Gidney2024cultivation,sunami2025entanglement}, improved decoding of concatenated codes~\cite{Gidney2025yoked}, and decoder switching~\cite{toshio2025decoder}. Since window decoding is a promising approach to real-time decoding, computing this quantity under window decoding is of practical importance. We note that the quantity discussed here differs in its target from the STCGs introduced in this work. The STCGs are designed to predict window-induced logical errors, whereas the conventional complementary gap quantifies the reliability of the logical class of the estimated error.

In global decoding, the conventional complementary gap is computed by first obtaining the minimum-weight error via MWPM, then flipping the boundary conditions on the two spatial boundaries and rerunning MWPM to obtain the minimum-weight error in the complementary logical class~\cite{Hutter2014,Bombin2024,Gidney2025yoked} (see also Sec.~\ref{subsec_comp_gap}). When one attempts to apply the same procedure in window decoding, two difficulties arise, analogous to those encountered in the computation of the STCG (see also Sec.~\ref{subsubsec_STCG}). First, if the virtual boundary is left open, it connects the two spatial boundaries, reducing the effective distance between them to a small constant independent of $d$. The resulting gap no longer reflects the reliability of the decoding result. Second, if the virtual boundary is simply closed, a defect that should have been matched to it must instead be matched to one of the spatial boundaries, which yields a spuriously small gap.

\begin{figure}
    \centering
    \includegraphics[width=\linewidth]{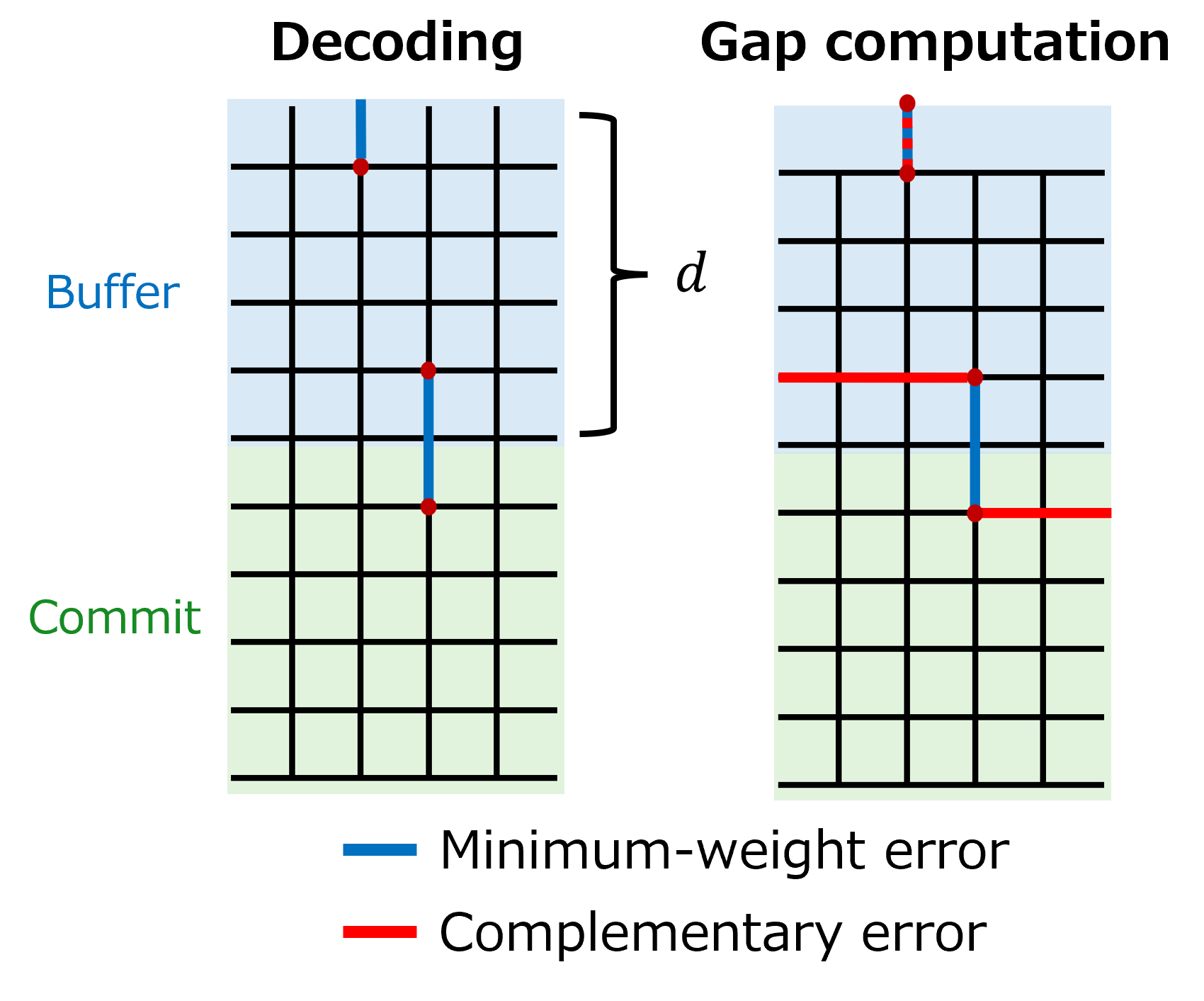}
    \caption{Procedure for computing the conventional complementary gap under sliding window decoding. After the minimum-weight error is first obtained in a window, the virtual boundary is modified, and the gap is then computed with the spatial boundary condition flipped.}
    \label{fig_cgap_computation}
\end{figure}

One can employ the same boundary modification as in the computation of the STCG to circumvent these difficulties; see Fig.~\ref{fig_cgap_computation}. After computing the minimum-weight error $E_\mathrm{min}$, the virtual boundary is closed, except for the edges included in $E_\mathrm{min}$, which are retained with the boundary node replaced by a new node at which an additional defect is placed. The complementary error $E_\mathrm{comp}$ is then computed with the  spatial boundary condition flipped. The gap for the entire decoding graph is taken to be the minimum of the gap values obtained for the individual windows, since the least reliable window dominates the reliability of the overall estimate.

In the procedure described above, the commit and buffer regions are not distinguished; the complementary gap is computed over the entire window. Since the gap ideally should reflect only the reliability of the commit region, a more precise computation would require restricting the contribution to the commit region, but how to do so is not straightforward. Nevertheless, the numerical results show that the proposed procedure already achieves sufficient precision, as demonstrated below.

\begin{figure*}
    \centering
  \begin{tabular}{ll}
{\normalsize (a)} & {\normalsize (b)}  \\
   \includegraphics[width=0.517\linewidth]{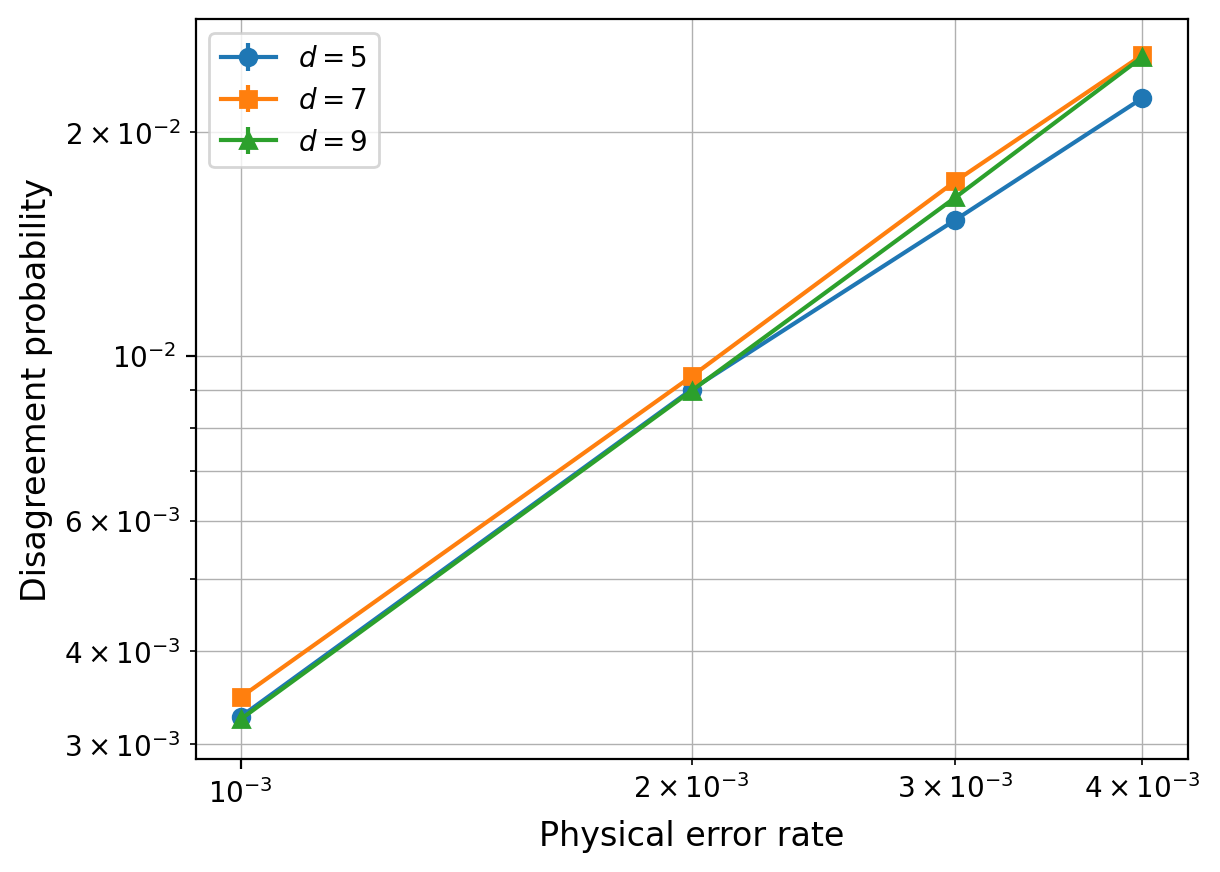}&
   \includegraphics[width=0.483\linewidth]{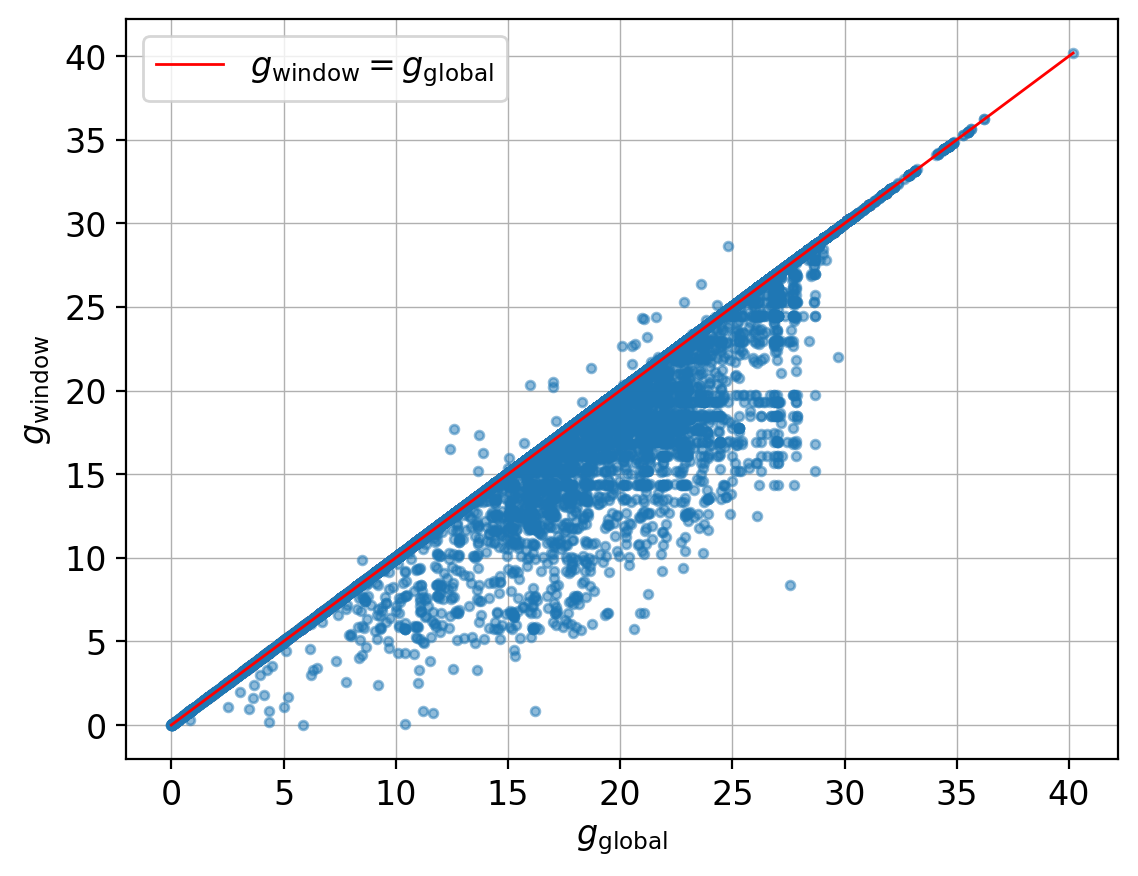}
   \\
  \end{tabular}
    \caption{Comparison of the complementary gap $g_{\mathrm{window}}$ computed under sliding window and $g_{\mathrm{global}}$ computed under global decoding, for the surface code under circuit-level noise. (a) Probability that $g_{\mathrm{window}}$ and $g_{\mathrm{global}}$ differ. Two gap values are regarded as different when their relative difference with respect to $g_{\mathrm{global}}$ is at least $\SI{1}{\%}$. (b) Scatter plot of $g_{\mathrm{window}}$ against $g_{\mathrm{global}}$ for $d=7$ and $p=0.001$. The total number of syndrome measurement rounds is $10d$. $r_{\mathrm{com}} = r_{\mathrm{buf}} = d$. The results are based on $10^6$ Monte Carlo samples, which are individually plotted in (b).}
    \label{fig_cgap_difference}
\end{figure*}

\begin{figure*}
    \centering
  \begin{tabular}{ll}
{\normalsize (a)} & {\normalsize (b)}  \\
   \includegraphics[width=0.5\linewidth]{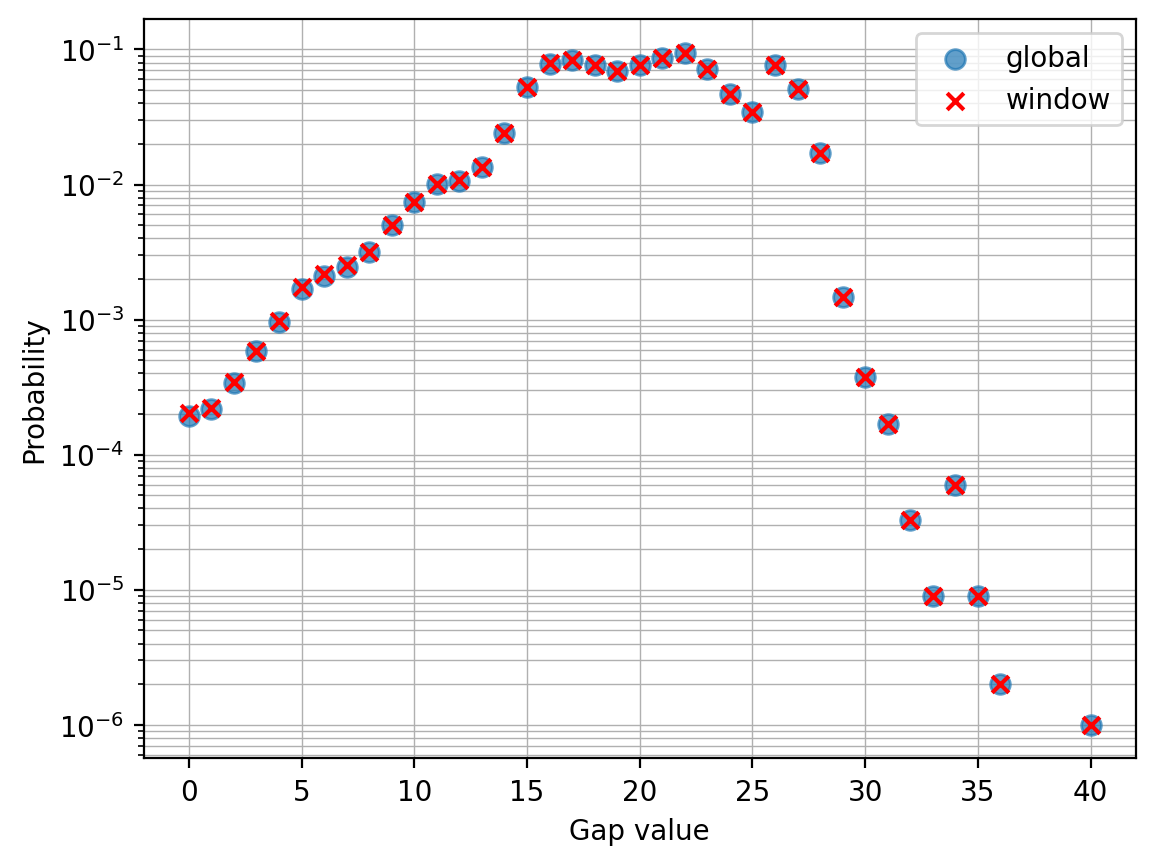}&
   \includegraphics[width=0.5\linewidth]{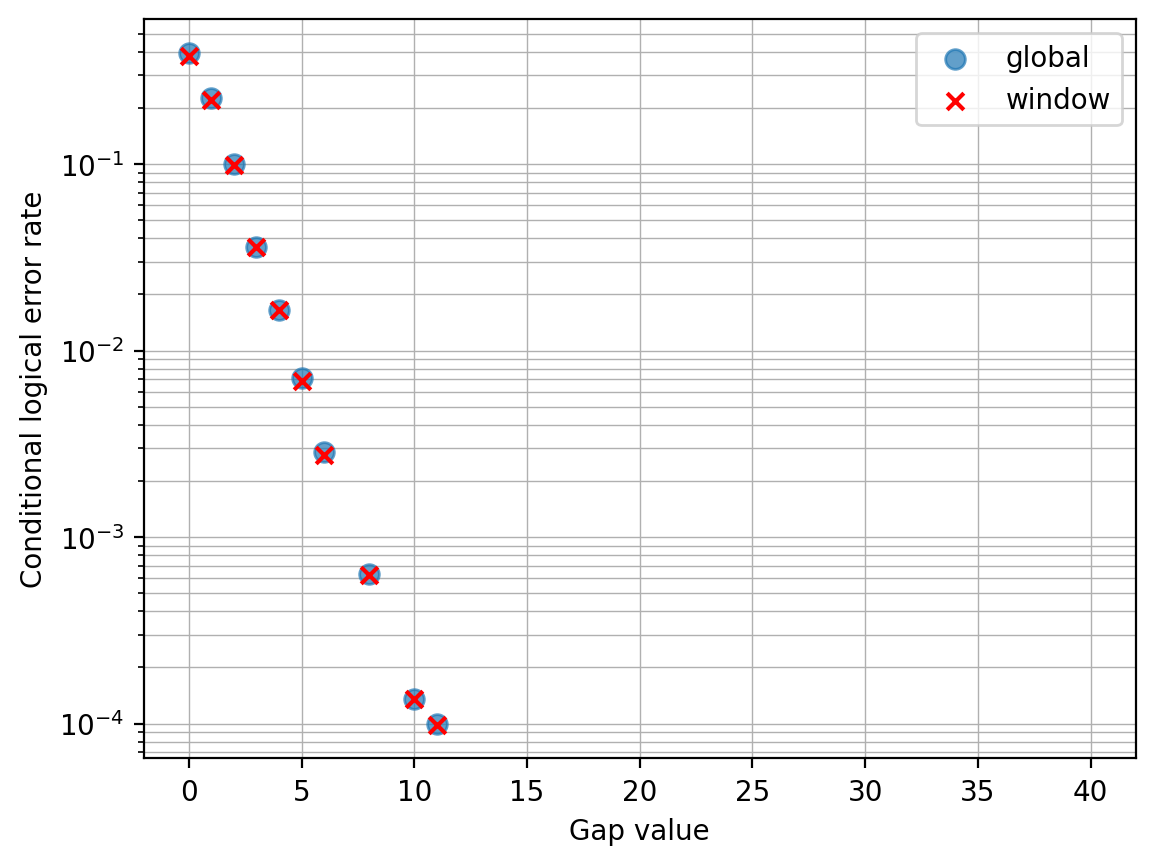}
   \\
  \end{tabular}
    \caption{Comparison of (a) the gap distribution and (b) the logical error rate conditioned on the gap value, computed under sliding window decoding and global decoding, for the surface code under circuit-level noise. The total number of syndrome measurement rounds is $10d$. $d=7$, $p=0.001$, and $r_{\mathrm{com}} = r_{\mathrm{buf}} = d$. The results are based on $10^6$ Monte Carlo samples.}
    \label{fig_cgap_property_comparison}
\end{figure*}

Figure~\ref{fig_cgap_difference} presents a comparison of the complementary gap $g_\mathrm{window}$ computed under sliding window decoding with the proposed method and $g_\mathrm{global}$ computed under global decoding, in a memory experiment on the surface code with circuit-level noise. Figure~\ref{fig_cgap_difference} (a) shows the probability that $g_\mathrm{window}$ and $g_\mathrm{global}$ differ. The probability shows no strong dependence on the code distance and remains small across the range of $d$ examined. It decreases further as the physical error rate is reduced. Figure~\ref{fig_cgap_difference} (b) is a scatter plot of $g_\mathrm{window}$ against $g_\mathrm{global}$. Most of the points that do not lie on the line $g_\mathrm{window} = g_\mathrm{global}$ are located in the region $g_\mathrm{window} < g_\mathrm{global}$. This indicates that, when the proposed method outputs an incorrect gap value, the value tends to be an underestimate. This means that the probability of incorrectly regarding an actually unreliable decoding result as reliable is small. This property is desirable for applications such as switching based on complementary gap.

Figure~\ref{fig_cgap_property_comparison} (a) compares the gap distributions and (b) compares the logical error rates conditioned on the gap value, computed under sliding window decoding and global decoding, respectively. Both quantities show close agreement between the two decoding schemes. This indicates that the complementary gap computed under sliding window decoding achieves sufficient accuracy for practical applications such as post-selection~\cite{Bombin2024,Gidney2024cultivation,sunami2025entanglement} and decoder switching~\cite{toshio2025decoder}.
\newpage

\clearpage

\bibliography{citation}

\end{document}